%% file: main.tex
\documentclass[sigconf, nonacm]{acmart}

\newcommand\vldbdoi{XX.XX/XXX.XX}
\newcommand\vldbpages{XXX-XXX}
\newcommand\vldbvolume{14}
\newcommand\vldbissue{1}
\newcommand\vldbyear{2020}
\newcommand\vldbauthors{\authors}
\newcommand\vldbtitle{\shorttitle} 
\newcommand\vldbavailabilityurl{https://cp.kaist.ac.kr/memento}
\newcommand\vldbpagestyle{plain} 

\input{setup}

\usepackage{tikz}
\usetikzlibrary{litmus, litmus.listings, decorations}
\litmusset{use listing={style=litmus, basicstyle=\footnotesize\ttfamily}}

\usepackage{tcolorbox}
\tcbuselibrary{listings}
\tcbuselibrary{skins}

\tcbset{
  listing options={
      language={[LaTeX]TeX},
      basicstyle=\footnotesize\ttfamily,
      commentstyle=\color{black!75},
      columns=flexible,
      keepspaces=true,
  },
  frame hidden,
  sharp corners,
  bicolor,
  top=0pt,
  bottom=0pt,
  left=0pt,
  right=0pt,
  colback=blue!10,
  colbacklower=white,
  sidebyside align=top seam,
}

\usepackage{hyperref}
\hypersetup{
    linkbordercolor={1 1 1},
    colorlinks=true,
    linkcolor=violet,
}

\begin{document}
\title{Practical Detectability for Persistent Lock-Free Data Structures}

\author{Kyeongmin Cho}
\affiliation{%
  \institution{KAIST}
  \streetaddress{Rm. 4441, Bldg. E3-1, KAIST, 291 Daehak-ro, Yuseong-gu}
  \city{Daejeon}
  \country{Korea}
  \postcode{34141}
}
\email{kyeongmin.cho@kaist.ac.kr}

\author{Seungmin Jeon}
\affiliation{%
  \institution{KAIST}
  \streetaddress{Rm. 4441, Bldg. E3-1, KAIST, 291 Daehak-ro, Yuseong-gu}
  \city{Daejeon}
  \country{Korea}
  \postcode{34141}
}
\email{seungmin.jeon@kaist.ac.kr}

\author{Jeehoon Kang}
\affiliation{%
  \institution{KAIST}
  \streetaddress{Rm. 4441, Bldg. E3-1, KAIST, 291 Daehak-ro, Yuseong-gu}
  \city{Daejeon}
  \country{Korea}
  \postcode{34141}
}
\email{jeehoon.kang@kaist.ac.kr}

\begin{abstract}
Persistent memory (PM) is an emerging class of storage technology that combines the benefits of DRAM and SSD.
This characteristic inspires research on persistent objects in PM with fine-grained concurrency control.
Among such objects, persistent lock-free data structures (DSs) are particularly interesting thanks to their efficiency and scalability.
One of the most widely used correctness criteria for persistent lock-free DSs is \emph{durable linearizability} (Izraelevitz \etal{}, DISC 2016).
However, durable linearizability is insufficient to use persistent DSs for fault-tolerant systems requiring exactly-once semantics for storage systems,
because we may not be able to detect whether an operation is performed when a crash occurs.

We present a practical programming framework for persistent lock-free DSs with detectability.
In contrast to the prior work on such DSs, our framework supports
\begin{enumerate*}
\item primitive detectable operations such as space-efficient compare-and-swap, insertion, and deletion;
\item systematic transformation of lock-free DSs in DRAM into those in PM requiring modest efforts;
\item comparable performance with non-detectable DSs by DRAM scratchpad optimization; and
\item recovery from both full system and thread crashes.
\end{enumerate*}
The key idea is \emph{memento} objects serving as a lightweight, precise, and per-thread checkpoints in PM.
As a case study, we implement lock-free and combining queues and hash tables with detectability that outperform (and perform comparably) the state-of-the-art DSs with (and without, respectively) detectability.
%
\end{abstract}

\maketitle

\pagestyle{\vldbpagestyle}
\begingroup\small\noindent\raggedright\textbf{PVLDB Reference Format:}\\
\vldbauthors. \vldbtitle. PVLDB, \vldbvolume(\vldbissue): \vldbpages, \vldbyear.\\
\href{https://doi.org/\vldbdoi}{doi:\vldbdoi}
\endgroup
\begingroup
\renewcommand\thefootnote{}\footnote{\noindent
This work is licensed under the Creative Commons BY-NC-ND 4.0 International License. Visit \url{https://creativecommons.org/licenses/by-nc-nd/4.0/} to view a copy of this license. For any use beyond those covered by this license, obtain permission by emailing \href{mailto:info@vldb.org}{info@vldb.org}. Copyright is held by the owner/author(s). Publication rights licensed to the VLDB Endowment. \\
\raggedright Proceedings of the VLDB Endowment, Vol. \vldbvolume, No. \vldbissue\ %
ISSN 2150-8097. \\
\href{https://doi.org/\vldbdoi}{doi:\vldbdoi} \\
}\addtocounter{footnote}{-1}\endgroup

\ifdefempty{\vldbavailabilityurl}{}{
\vspace{.3cm}
\begingroup\small\noindent\raggedright\textbf{PVLDB Artifact Availability:}\\
The source code, data, and/or other artifacts have been made available at \url{\vldbavailabilityurl}.
\endgroup
}

{\input{intro}}
{\input{overview}}
{\input{cas}}
{\input{smo}}
{\input{volatile}}
{\input{reclamation}}
{\input{evaluation}}
{\input{conclusion}}



\bibliographystyle{ACM-Reference-Format}
\bibliography{reference}

\end{document}

%% file: setup.tex
\usepackage{hyperref}
\usepackage[capitalise,nameinlink]{cleveref}
\usepackage{url}
\usepackage{datetime}
\usepackage{xspace}
\usepackage{wasysym}
\usepackage{mathpartir}
\usepackage{dashbox}
\usepackage[inline]{enumitem}
\setlist{leftmargin=4mm,topsep=2mm}
\setlist[enumerate]{label=\textbf{(\arabic*)}}
\usepackage{colortbl}
\usepackage{tikz}
\usepackage{multirow}
\usepackage{booktabs}
\newlist{tabitemize}{itemize}{1}
\setlist[tabitemize]{leftmargin=2.5mm,label=\textbullet,nosep,after=\strut}
\usepackage{tabularx}
\usepackage[inkscapearea=page]{svg}
\usepackage{circledsteps}
\usepackage{subcaption}
\usepackage{algorithm}
\usepackage{algpseudocode}

\usepackage{pifont}

\usepackage{upquote}
\usepackage{kotex}

\newtheorem{reqthm}{Requirement}

\newcounter{mylabelcounter}

\makeatletter
\newcommand{\labelAxiom}[2]{%
\hfill{\textsc{(#1)}}\refstepcounter{mylabelcounter}
\immediate\write\@auxout{%
  \string\newlabel{#2}{{\unexpanded{\textsc{#1}}}{\thepage}{{\unexpanded{\textsc{#1}}}}{mylabelcounter.\number\value{mylabelcounter}}{}}
}%
}
\makeatother

\definecolor{StringRed}{rgb}{.637,0.082,0.082}
\definecolor{CommentGreen}{rgb}{0.0,0.55,0.3}
\definecolor{KeywordBlue}{rgb}{0.0,0.3,0.55}
\definecolor{LinkColor}{rgb}{0.55,0.0,0.3}
\definecolor{CiteColor}{rgb}{0.55,0.0,0.3}
\definecolor{HighlightColor}{rgb}{0.0,0.0,0.0}

\definecolor{grey}{rgb}{0.5,0.5,0.5}
\definecolor{red}{rgb}{1,0,0}
\definecolor{darkgreen}{rgb}{0.0,0.7,0.0}

\hypersetup{%
  linktocpage=true, pdfstartview=FitV,
  breaklinks=true, pageanchor=true, pdfpagemode=UseOutlines,
  plainpages=false, bookmarksnumbered, bookmarksopen=true, bookmarksopenlevel=3,
  hypertexnames=true, pdfhighlight=/O,
  colorlinks=true,linkcolor=LinkColor,citecolor=CiteColor,
  urlcolor=LinkColor
}

\crefdefaultlabelformat{\textcolor{ACMPurple}{#2#1}#3}

\crefname{section}{\textcolor{ACMPurple}{\S{}}}{\textcolor{ACMPurple}{\S{}}\!}
\crefformat{section}{\textcolor{ACMPurple}{#2\S{}#1#3}}
\Crefname{section}{Section}{Sections}
\Crefformat{section}{\textcolor{ACMPurple}{\S{}#2#1#3}}
\crefrangelabelformat{section}{\textcolor{ACMPurple}{#3#1#4}--\textcolor{ACMPurple}{#5#2#6}}

\Crefname{figure}{\textcolor{ACMPurple}{\text{Figure}}}{\textcolor{ACMPurple}{\text{Figures}}}
\crefname{figure}{\textcolor{ACMPurple}{\text{Figure}}}{\textcolor{ACMPurple}{\text{Figures}}}
\Crefname{table}{\textcolor{ACMPurple}{\text{Table}}}{\textcolor{ACMPurple}{\text{Tables}}}
\crefname{table}{\textcolor{ACMPurple}{\text{Table}}}{\textcolor{ACMPurple}{\text{Tables}}}
\Crefname{algorithm}{\textcolor{ACMPurple}{\text{Algorithm}}}{\textcolor{ACMPurple}{\text{Algorithms}}}
\crefname{algorithm}{\textcolor{ACMPurple}{\text{Algorithm}}}{\textcolor{ACMPurple}{\text{Algorithms}}}
\crefname{corollary}{\text{Corollary}}{\text{corollaries}}
\Crefname{corollary}{\text{Corollary}}{\text{Corollaries}}
\crefname{lemma}{\text{Lemma}}{\text{Lemmas}}
\Crefname{lemma}{\text{Lemma}}{\text{Lemmas}}
\crefname{theorem}{\text{Theorem}}{\text{Theorems}}
\Crefname{theorem}{\text{Theorem}}{\text{Theorems}}
\crefname{proposition}{\text{Prop.}}{\text{Propositions}}
\Crefname{proposition}{\text{Proposition}}{\text{Propositions}}
\crefname{definition}{\text{Def.}}{\text{Definitions}}
\Crefname{definition}{\text{Definition}}{\text{Definitions}}
\Crefname{reqthm}{\textcolor{ACMPurple}{\text{Requirement}}}{\textcolor{ACMPurple}{\text{Requirements}}}
\crefname{reqthm}{\textcolor{ACMPurple}{\text{Requirement}}}{\textcolor{ACMPurple}{\text{Requirements}}}

\algrenewcommand\algorithmicindent{1.3em}
\algnewcommand\algorithmicswitch{\textbf{switch}}
\algnewcommand\algorithmiccase{\textbf{case}}
\algnewcommand\algorithmicassert{\texttt{assert}}
\algnewcommand\Assert[1]{\State \algorithmicassert(#1)}%
\algnewcommand{\IfThenElse}[3]{
  \State \algorithmicif\ #1\ \algorithmicthen\ #2\ \algorithmicelse\ #3}
\algnewcommand{\IfThen}[2]{
  \State \algorithmicif\ #1\ \algorithmicthen\ #2}
\algdef{SE}[SWITCH]{Switch}{EndSwitch}[1]{\algorithmicswitch\ #1\ \algorithmicdo}{\algorithmicend\ \algorithmicswitch}%
\algdef{SE}[CASE]{Case}{EndCase}[1]{\algorithmiccase\ #1}{\algorithmicend\ \algorithmiccase}%
\algtext*{EndSwitch}%
\algtext*{EndCase}%

\renewcommand{\paragraph}[1]{\noindentparagraph{\textbf{#1}.}}

\newcommand{\code}[1]{\texttt{#1}}

\newcommand{\eg}{\emph{e.g.}}
\newcommand{\ie}{\emph{i.e.}}
\newcommand{\etal}{\emph{et.~al.}}


%% file: intro.tex
\section{Introduction}

Persistent memory (PM) is an emerging class of storage technology that
simultaneously provides
\begin{enumerate*}
	\item low latency, high throughput, and fine-grained data transfer unit as DRAM does; and
	\item durability and high capacity as SSD does.
\end{enumerate*}
Compared with DRAM, Intel Optane DC Persistent memory module (DCPMM)~\cite{optane-dc} has 3$\times$ and similar latency for read and write; $1/3$ and $1/6$ bandwidth for read and write~\cite{nvm-performance}; and up to 4$\times$ capacity for a single DIMM slot.
Thanks to these characteristics, PM has a great potential for optimizing traditional and distributed file systems~\cite{strata,winefs,kuco,octopus,linefs,nova}, transaction processing systems for high-velocity real-time data~\cite{sstore}, distributed stream processing systems~\cite{kafka}, and stateful applications organized as a pipeline of cloud serverless functions interacting with cloud storage systems~\cite{olive,beldi}.


A key ingredient of such optimization is persistent \emph{lock-free} data structures (DSs), which provide the following advantages over lock- and transaction-based ones.
%
\emph{First}, they significantly outperform lock- and transaction-based DSs, especially in the presence of contention~\cite{link-persist}, by allowing multiple threads to process queries in parallel just as in DRAM.
For instance, lock-free FIFO queues significantly outperform lock- and transaction-protected ones, as we will see in \cref{sec:evaluation}.
%
%
\emph{Second}, lock-free algorithms ensure that DSs are always in a consistent state so that we do not need additional logging to ensure crash consistency, as far as the writes to DSs are flushed to PM in a timely manner.
For these reasons, persistent lock-free DSs have drawn significant attention in the literature.
On the one hand, several such DSs as stacks~\cite{tracking}, queues~\cite{friedman}, lists~\cite{tracking, soft}, hash tables~\cite{clevel, soft, cceh}, and trees~\cite{tracking}, have been proposed.
On the other hand, several general transformations from lock-free DSs in DRAM into persistent DSs in PM have been proposed~\cite{recipe,nvtraverse,mirror,izraelevitz}.


One of the most widely accepted correctness criteria for persistent lock-free DSs is \emph{durable linearizability}~\cite{izraelevitz}.
For lock-free DSs in DRAM, an operation's execution is modeled as an interval between its \emph{invocation} and \emph{response} events, and a thread's execution is modeled as a series of consecutive invocation and response events of its operations.
Then an execution of multiple threads are called \emph{linearizable} if all operations result in outputs \emph{as if} each of them atomically takes effect at some point, called the \emph{linearization point}, between its invocation and response~\cite{linearizability}.
Now for lock-free persistent DSs in PM, an execution possibly across multiple crashes is called \emph{durably linearizable} if it is linearizable when ignoring the crash events.
In particular, operations finished before a crash should be persisted across the crash (durability), and even if some operations are interrupted in the middle by a crash, the DS should be \emph{recovered} to a consistent state right after the crash (consistency).
Finally, a persistent lock-free DS is durably linearizable if so are all its executions~\cite{izraelevitz}.
Durable linearizability is indeed satisfied by most existing persistent lock-free DSs, except for those DSs that intentionally break the property in order to trade durability for performance~\cite{friedman}. 

\begin{figure}
  \centering
  \includesvg[width=\columnwidth]{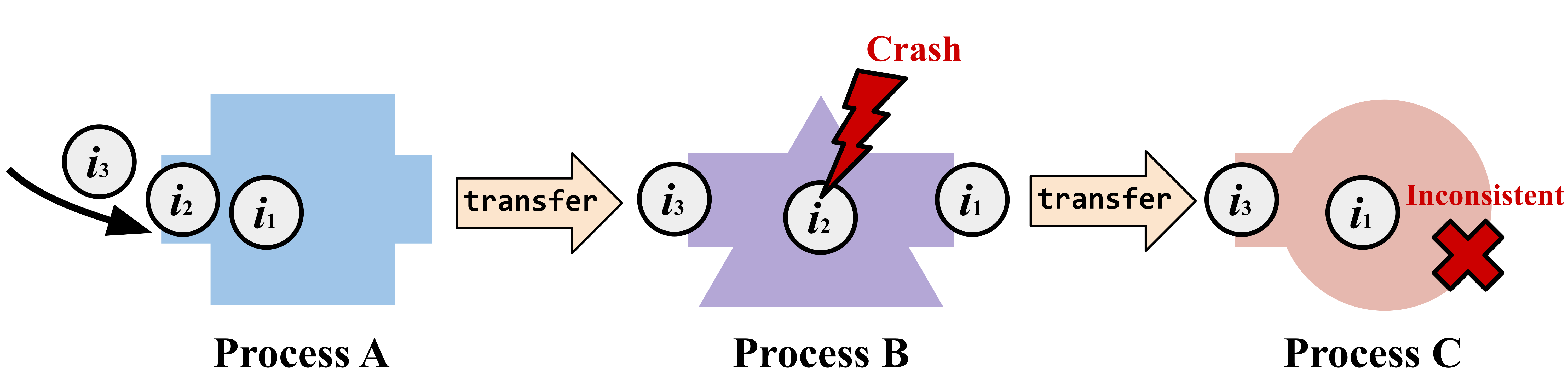}
  \vspace{-2em}
  \caption{Transaction Processing System with Persistent DSs}
  \label{fig:intro:correctness}
  \vspace{-0.7em}
\end{figure}

We argue that durable linearizability, while being widely accepted, is generally insufficient to use persistent lock-free DSs for transaction processing.
Durable linearizability does not tell about \emph{detectability}, or in other words, when an operation is interrupted in the middle by a crash, whether it is finished or not~\cite{friedman,nrl,nrlp,dss}.
The lack of detectability---while ensuring the DS's own consistency---may break the enclosing system's consistency in case of crashes.
\cref{fig:intro:correctness} illustrates a transaction processing system utilizing a persistent lock-free DS for high-velocity real-time data~\cite{sstore,winter}.
%
In such a system, a component DS consumes and produces a stream of inputs and outputs, respectively~\cite{kafka}.
Suppose the DS is not detectable.
Should the system crashes, we cannot detect whether the input being processed just before the crash, say $i_2$, is finished, possibly incurring an inconsistency between the DS and the rest of the system and thus unrecoverable errors.
For instance, If the system were a banking system and the DS were a key-value store recording the amount of balance for each person, a crash in the middle of a deposit operation may lose a person's balance forever;
for another instance, if the system were a distributed file system and the DS were a log, a storage node's crash may silently ignore an ongoing operation, incurring inconsistency across the nodes.







\paragraph{Challenges}


Several detectable persistent lock-free DSs have been proposed in the literature~\cite{friedman,dss,rusanovsky,nrl,nrlp}, but to the best of our knowledge, all of them suffer from at least one of the following limitations:


\begin{itemize}
\item \emph{High design cost}: The prior works present several such persistent lock-free DSs, but many of them are hand-tuned and are manually reasoned to ensure crash consistency and detectability:
  Friedman \etal~\cite{friedman} and Li \etal{}~\cite{dss} present detectable persistent lock-free queues, and Rusanovsky \etal{}~\cite{rusanovsky} and Fatourou \etal{}~\cite{pcomb} present detectable persistent combining DSs.
  Attiya \etal{}~\cite{nrl} present detectable compare-and-swap (CAS) for PM location as a general primitive operation for pointer-based DSs,
  but its applicability to lock-free DSs have not been demonstrated.
  Ben-David \etal{}~\cite{nrlp} presents a transformation based on detectable CAS from lock-free DSs in DRAM into those in PM with detectability,
  but it requires code to follow specific patterns such as the \emph{normalized} form~\cite{normalized}; and sizable restructuring of code by inserting persistency boundaries and explicit recovery codes.

  %
\item \emph{High runtime overhead}: Detectability generally incurs runtime overhead.
  While the overhead is modest for hand-tuned persistent DSs~\cite{friedman,dss,rusanovsky}, it is significant for the general transformation in \cite{nrlp} for two reasons.
  \emph{First}, an object supporting detectable CAS consumes $O(P)$ space in PM where $P$ is the number of threads, prohibiting its use for space-efficient DSs such as hash tables and trees.
  (More severely, Attiya \etal{}~\cite{nrl}'s detectable CAS object consumes $O(P^2)$ space in PM.)
  \emph{Second}, the transformation needs to checkpoint in PM stack variables for each CAS unless it is optimized for normalized DSs~\cite{normalized}.
\item \emph{Unsafe memory reclamation}:
  Friedman \etal~\cite{friedman} use hazard pointers~\cite{hp}, and Li \etal{}~\cite{dss} and (nb)Montage~\cite{montage,nbmontage} use epoch-based reclamation~\cite{ebr} for reclaiming PM locations, but its technical details are not sufficiently discussed in the papers.
  %
  Attiya \etal{}~\cite{nrl}, Ben-David \etal{}~\cite{nrlp}, and Rusanovsky \etal{}~\cite{rusanovsky}
  Other prior works on (detectable or non-detectable) persistent lock-free DSs in PM~\cite{nrl,nrlp,rusanovsky} do not discuss safe memory reclamation.
  We argue that reclamation for PM is extremely subtle and deserves more attention.
  \emph{First}, we discover that the queues in \cite{friedman,dss} may incur use-after-free in case of crashes due to a lack of flush.
  \emph{Second}, a straightforward application of reclamation schemes to PM may incur double-free in case of thread crashes.
\end{itemize}

\paragraph{Contributions and Outline}

In this work, we present a programming framework for persistent lock-free DSs with detectability in PM that overcomes the above limitations.
Specifically, we make the following contributions:

\begin{itemize}
\item In \cref{sec:overview}, we propose a model of detectable operations supporting systematic and efficient composition with end-to-end exactly-once semantics.
The key idea is persisting not only DS but also per-thread, timestamp-based \emph{memento} objects serving as a precise, lightweight, and mostly thread-local checkpoints for the corresponding thread's ongoing operation on the DS.


\item In \cref{sec:cas}, we design an atomic pointer location object supporting efficient detectable CAS consuming $O(1)$ space in PM.
Our CAS object is based on those of Attiya \etal{}~\cite{nrl} and Ben-David \etal{}~\cite{nrlp}, which utilize \emph{per-object} array of $O(P^2)$ and $O(P)$ sequence numbers, respectively.
We reduce the space complexity by instead utilizing \emph{global} arrays of $O(P)$ timestamps.

\item In \cref{sec:smo}, we design detectable insertion and deletion operations.
While CAS is a general primitive operation for pointer-based DSs, we observe CAS-based detectable DSs are sometimes significantly slower than hand-tuned detectable DSs due to the increased contention on the CAS location.
As an optimization, we design a more efficient (and less general) detectable insertion and deletion operations by capturing the essence of the hand-tuned detectable DSs presented in \cite{friedman,dss}.

\item In \cref{sec:volatile}, we propose a guideline for optimizing detectable DSs with DRAM scratchpad while preserving detectability.
By DRAM scratchpad we mean putting a part of persistent DSs in DRAM.
Such an optimization is pioneered in SOFT~\cite{soft} and Mirror~\cite{mirror} that replicate PM contents in DRAM for higher performance, but it is neither supported by our composition (\cref{sec:overview}) nor explicitly recognized and exploited for detectable DSs in the literature.
We explain how to add a DRAM scratchpad to a detectable DS while preserving its detectability.


\item In \cref{sec:reclamation}, we propose correctness criteria for safe memory reclamation of lock-free DSs in PM.
We explain a use-after-free bug in \cite{friedman,dss} and two potential errors when reclamation schemes are straightforwardly applied in case of thread crashes.
We fix the bug and prevent the errors by generalizing reclamation schemes and clearing mementos.


\item In \cref{sec:evaluation}, we implement and evaluate detectable persistent lock-free DSs on Intel Optane DCPMM.
We implement detectable versions of MSQ, combining queue, Treiber's stack~\cite{treiber}, exchanger~\cite{}, elimination stack~\cite{elim}, Clevel hash table~\cite{clevel}, and SOFT hash table~\cite{soft} based on mementos.
Our MSQ outperforms the existing persistent MSQs with and without detectability; and
our combining queue, Clevel, and SOFT performs comparably with (and incurs a slight overhead over) the existing persistent DSs without detectibability for read-heavy (and write-heavy, respectively) workloads.

%

\end{itemize}


\noindent To summarize, we present the first programming framework for detectable persistent lock-free DSs with low design cost, modest runtime overhead, and safe memory reclamation at the same time.
Our implementation and evaluation result are available as supplementary material and they will be open-sourced after review.





%% file: overview.tex
\section{Design of Detectable Operations}
\label{sec:overview}

\subsection{Overview}
\label{sec:overview:overview}

\begin{figure}
  \centering
  \includesvg[width=\columnwidth]{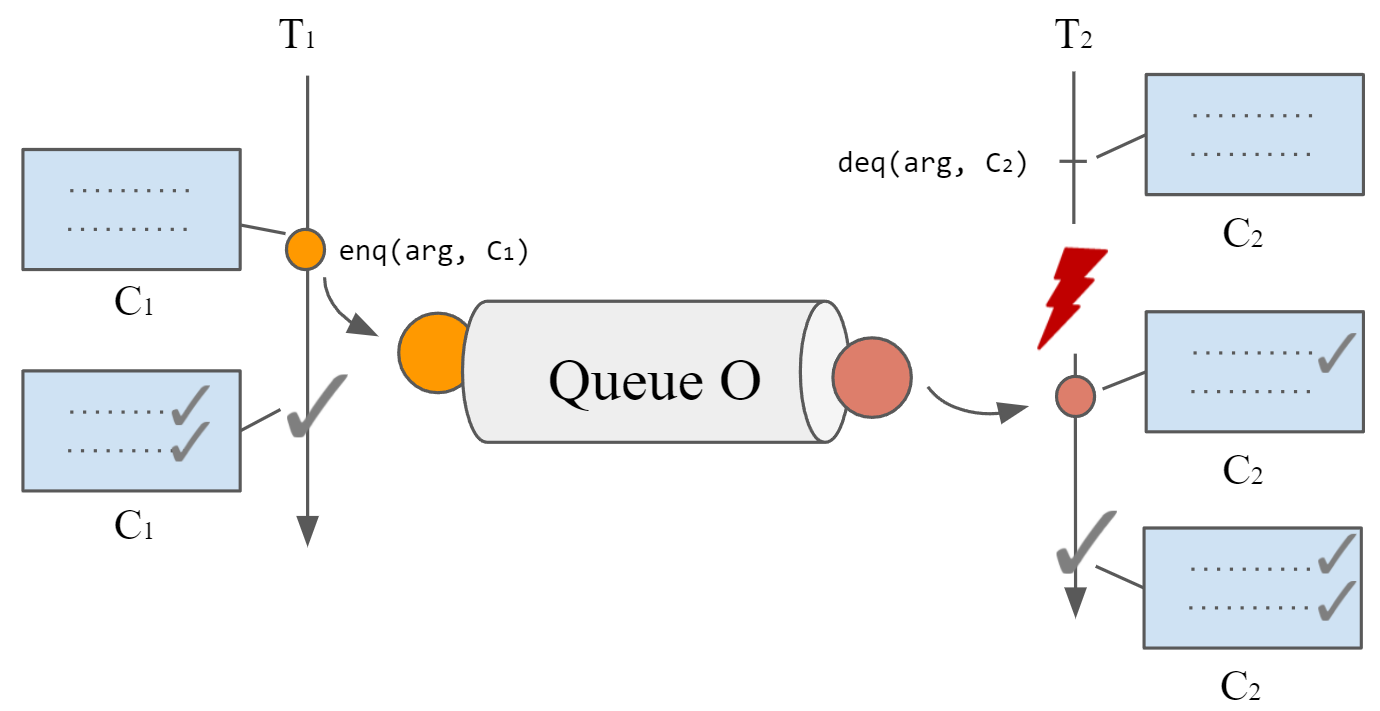}
  \vspace{-2em}
  \caption{A Queue and its Mementos}
  \label{fig:framework:overview}
  \vspace{-0.7em}
\end{figure}


We design detectable persistent lock-free DSs in PM using per-thread mementos.
\cref{fig:framework:overview} illustrates an example object $O$ in PM that is concurrently accessed by threads $T_1$ and $T_2$.
Suppose $O$ is a queue and $T_1$ and $T_2$ are performing enqueue and dequeue operations, respectively.
We require $O$'s operations receive---in addition to ordinary arguments such as the value to enqueue---a reference to memento stored in PM that is used to checkpoint the operation's progress.
Each memento supports a specific type of operations, \eg{}, the mementos $C_1$ and $C_2$ for $T_1$ and $T_2$ support enqueue and dequeue operations, respectively.
The given memento records how far the operation had made the progress.
In particular, once an operation is finished, the given memento records the operation's output so that it can later be retrieved.
%

Even when interrupted in the middle by crashes, a thread still ensures its operation's exactly-once semantics by detecting the progress it has made from the memento.
Specifically, should the system or a thread crash, a crashed thread simply re-executes its operation and then it will automatically
\begin{enumerate*}
\item detect the progress it has made before the crash and replay its effects; and
\item perform the remaining tasks and checkpoint the further progress in the memento.
\end{enumerate*}
As a result, an operation makes its effect exactly once---neither never nor twice---even in case of multiple crashes; and returns the same output before and after crashes.
%


%
%


%
%

For correct checkpointing of the progress, a memento should record in PM exactly which operation steps are already performed and what are their outputs.
In general, we construct such a memento by \emph{composing primitive operations}.
\emph{First}, we design mementos for primitive operations such as
\begin{enumerate*}
\item checkpointing of non-deterministic computations; and
\item compare-and-swap, insertion, and deletion of memory blocks.
\end{enumerate*}
The former is necessary to ensure the operation produces the same output after crashes; and
the latter serves as the basis for pointer-based lock-free DSs.
%
%
%
%
\emph{Second}, we compose multiple mementos to construct that for a composite operation while retaining detectability.
As we will see shortly, we support sequence, conditional branch, and loop compositions of operations.

%







\subsection{Primitive Operation Example: Checkpoint}
\label{sec:overview:example}

As an example of an operation with memento, we present the \textsc{Checkpoint} operation recording a value so that it is preserved after crashes.
This operation effectively \emph{stabilizes} possibly non-deterministic input to a deterministic output so that, \eg{}, the operation always takes the same control flow across crashes (see \cref{sec:overview:stability} for details on the requirement for correct recovery).
You may checkpoint an expression as follows:


%
\begin{algorithmic}[1]
  \State{$v \gets \textsc{Checkpoint}(e, mmt)$}
\end{algorithmic}
Let $e$ be a non-deterministic expression and $mmt$ be a checkpoint variable memento.
Then the value $v$ is preserved across crashes.

\paragraph{Algorithm}

We ensure \textsc{Checkpoint}'s atomicity---we can never see a partial result of the computation---by defining a memento as two timestamps and two write buffers.
An operation
\begin{enumerate*}
\item[\ding{182}] compares the given memento's two timestamps and figure out which are stale and latest;
\item[\ding{183}] if the latest timestamp is bigger than those of the thread's previous operations, exits immediately;
\item[\ding{184}] writes the given value to the memento's stale buffer;
\item[\ding{185}] flushes the stale buffer if the memento is not fit in a cacheline; and
\item[\ding{186}] updates the stale buffer's timestamp to the current timestamp and flushes it.
\end{enumerate*}
Here, \ding{183} detects whether the operation is already performed in a pre-crash execution; and
the flush in \ding{185} ordering the writes to buffer and to timestamp, is unnecessary if they are fit in a single cacheline because all writes to a single cacheline in PM are automatically ordered~\cite{building-blocks}.


\paragraph{System Assumptions}

We retrieve timestamps by Intel-x86's \code{rdtsc} and \code{rdtscp} instructions.
We use Intel-x86's \code{flush} and \code{flushopt} instructions to ensure a write to a PM location is persisted.
More specifically, a store or update to a PM location, say $loc$, is guaranteed to be persisted
\begin{enumerate*}
\item at a following $\code{flush}~loc$ instruction; or
\item at a following \code{sfence} or successful CAS instruction preceded by a $\code{flushopt}~loc$ instruction.
\end{enumerate*}
We refer to \cite{pview} for more details on the semantics of \code{flush} and \code{flushopt} for Intel Optane DCPMM.


\paragraph{Comparison}

Our \textsc{Checkpoint} algorithm differs from how \cite{nrlp} persists stack variables with two buffers and a valid bit in that ours additionally records the timestamp when the value is checkpointed.
The timestamps enable each operation to detect, at recovery, whether it has already been executed before the crash.
For instance, consider the following $\textsc{Contrived}(\underline{i_1}, \underline{i_2}, mmt)$ operation:
\begin{algorithmic}[1]
  \State{$\underline{v_1} \gets \textsc{Checkpoint}(e_1, mmt.cv_1)$} \Comment{timestamp: 10}
  \Loop
  \State{$\underline{v_2} \gets \textsc{Checkpoint}(e_2, mmt.cv_2)$} \Comment{timestamp: 40}
  \State{$\underline{v_3} \gets \textsc{Checkpoint}(e_3, mmt.cv_3)$} \Comment{timestamp: 20}
  \State{$\underline{v_4} \gets \textsc{Cas}(\underline{v_2},\underline{v_3}, mmt.cas_4)$} \Comment{timestamp: 30}
  \IfThen{$\underline{v_1 = (v_4 ~?~ i_1 : i_2)}$}{$\textbf{return}$~$\underline{v_4}$}
  \EndLoop
\end{algorithmic}
The operation is given two inputs $i_1$ and $i_2$ and a memento $mmt$, and returns a value as an output at line 6 (or L6 from now on).
we will discuss the meaning of \underline{underlines} shortly.
The sub-memento $mmt.f$ for each $f$ is the $f$ field of the whole operation's composite memento $mmt$.
Suppose this operation is interrupted by a crash between L3 and L4 after a few iterations of the loop, checkpointing timestamps as commented.
At a recovery, the operation executes the code from the beginning and retrieves the checkpointed values at L1 and L3 as their timestamp increases, but checkpoints a new value at L4 and resumes from there as a normal execution because the timestamp no longer increases.
We record the maximum timestamp of each thread's previous operations in the global array $LOCAL$ in DRAM.
The array is used for other detectable operations as well (see \cref{sec:cas} for details).

Thanks to timestamps, our algorithm flushes fewer writes to PM than prior works: \cite{nrlp} needs to additionally checkpoint the program counter in PM, and \cite{nrl,dss} need to reset operations by writing sentinel values to PM.

\subsection{Stable Composition}
\label{sec:overview:stability}

For correct checkpointing of an operation's progress, we need not only to use primitive detectable operations but also to stably compose them as follows.
For simplicity, we assume the operation is in the Static Single Assignment (SSA) form~\cite{ssa1,ssa2} so that each local variable is defined only once syntactically, and loop variables are defined as an SSA $\phi$-node at the beginning of a basic block, \eg{}, $i = \phi(i_0 @ B_0, i_1 @ B_1)$ assigns $i_0$ (or $i_1$) to $i$ if the control flows from the basic block $B_0$ (or $B_1$, respectively).
For those who are not familiar with SSA, we will present an SSA form example shortly.


\begin{reqthm}[Stable memento]
  Different (syntactic) sub-operations of an operation are performed with different sub-mementos of the whole operation's memento.
  \label{req:overview:memento}
\end{reqthm}
\noindent
As a result, the sub-mementos are not interfered with each other.
%
For instance, \textsc{Contrived} satisfies this requirement as its sub-operations at L1, L3, L4, L5 use different sub-mementos of $mmt$.

\begin{reqthm}[Stable value]
  Branch conditions and \emph{certain} sub-operation inputs are \emph{stable}, \ie{}, they are the same across crashes.
  \label{req:overview:value}
\end{reqthm}
\noindent
As a result, an operation takes the same control flows and its sub-operations produce the same outputs across crashes.
\noindent Throughout this paper, we \underline{underline} stable expressions.
For instance, \textsc{Contrived} satisfies this requirement because the branch condition $(v_1 = (v_4 ~?~ i_1 : i_2))$ at L6 is stable as $v_1$ and $v_4$ are sub-operation outputs and $i_1$ and $i_2$ are stable inputs; and
the inputs $v_2$ and $v_3$ at L5 are stabilized by \textsc{Checkpoint}.

%
Here, we allow each operation to declare which inputs should be stable, \eg{},
\begin{enumerate*}
\item the expression input to \textsc{Checkpoint} (\eg{}, $e_1$ at L1) does \emph{not} need to be stable; while
\item the current and new pointer value inputs to detectable \textsc{Cas} (\eg{}, $v_2$ and $v_3$ at L5) should be stable.
\end{enumerate*}
This flexible requirement for input stability enables optimized implementation of the operations (see \cref{sec:smo} for details).



%

\begin{reqthm}[Stable loop]
  Suppose a loop contains at least one detectable sub-operation using memento. Then
  \begin{enumerate*}
  \item the loop head's basic block has at most one $\phi$-node; and
  \item the $\phi$-node, if exists, is immediately checkpointed.
  \end{enumerate*}
  \label{req:overview:loop}
\end{reqthm}
\noindent
Loops are interesting because loop bodies may reuse mementos despite \cref{req:overview:memento}, \eg{}, $mmt.cv_2$ is used to checkpoint different values in different iterations of the loop at L2-L7.
This requirement ensures the loop body's sub-operation results of two consecutive iterations are not mixed in the sub-mementos.
%
%
For instance, \textsc{Contrived} trivially satisfies this requirement as its the only loop has no $\phi$-nodes.
For another instance, consider the following operation:
\begin{algorithmic}[1]
  \State{$i_0 \gets e$}
  \Loop
  \State{$\underline{i} \gets \textsc{Checkpoint}(\phi(i_0, i_{next}), mmt)$}
  \IfThen{$\underline{i = 100}$}{$\textbf{break}$}
  \State{$i_{next} \gets i + 1$}
  \EndLoop
\end{algorithmic}
\noindent
Here, we omit the $\phi$-node's incoming blocks because they are obvious from the context.
The semantics of $\phi(i_0, i_{next})$ of the loop is that it gets $i_0$ for first iteration and $i_{next}$ for the remaining iterations.
This example also satisfies the requirement as the loop has a single $\phi$-node that is immediately checkpointed at L3.

The checkpointed $\phi$-node serves as the context for a loop iteration.
The sub-requirement \textbf{(2)} says the context should be recorded for each iteration; and \textbf{(1)} says the context should be \emph{atomically} recorded so that it is not mixed across multiple iterations.
Together, the requirement effectively guarantees you can correctly recover the loop iteration context, and from there, the progress that has been made so far inside the loop.

\subsection{Transformation}
\label{sec:overview:transformation}

Given a lock-free DS operation for DRAM, we systematically transform it into a detectable operation for PM by stably composing its sub-operations as follows.
For instance, \textsc{Contrived} can be transformed from the below code as follows:
\begin{algorithmic}[1]
  \State{$v_1 \gets e_1$}
  \Loop
  \State{$v_4 \gets \textsc{CasPlain}(e_2,e_3)$} \Comment{CAS without detectability}
  \IfThen{$v_1 = (v_4 ~?~ i_1 ~:~ i_2)$}{$\textbf{return}$~$v_4$}
  \EndLoop
\end{algorithmic}
%
%
\begin{enumerate*}
\item[\ding{182}] We replace plain CASes with detectable CASes (\cref{sec:cas}) or insert/delete operations (\cref{sec:smo}).
  We use a fresh sub-memento for each detectable CAS or insertion/deletion to satisfy \cref{req:overview:memento}.
  For the example, we choose to replace \textsc{CasPlain} with \textsc{Cas}, introducing a fresh sub-memento for the sub-operation.

\item[\ding{183}] We perform the SSA transformation~\cite{ssa1,ssa2}
  Further, for each loop containing at least one detectable operation, we merge all $\phi$-nodes into a single $\phi$-node of tuples and checkpoint it to satisfy \cref{req:overview:loop}.
  For the example, we have nothing to do because it does not need $\phi$-nodes.

\item[\ding{184}] We insert \textsc{Checkpoint} operations using fresh sub-mementos to satisfy \cref{req:overview:value}.
  We choose expressions to checkpoint using backward taint analysis from branch conditions and sub-operation inputs.
For the example, we stabilize $e_2$, $e_3$, and $(v_1 = (v_4 ~?~ i_1 ~:~ i_2))$ by checkpointing $e_2$, $e_3$, and $v_1$.
We do not need to checkpoint the other variables because $v_4$ is the stable result of a detectable CAS and we assume $i_1$ and $i_2$ are stable inputs.
We could alternatively checkpoint $(v_1 = (v_4 ~?~ i_1 ~:~ i_2))$ as a whole; or checkpoint $i_1$ and $i_2$ and assume they are unstable inputs.
We heuristically choose as above to minimize the number of \textsc{Checkpoint} operations.


\end{enumerate*}


\subsection{Top-Level Application}
\label{sec:overview:application}

At the top level, an application is structured as a root object and its mementos persisted in PM.
The monitor manages the execution of an application by
\begin{enumerate*}
\item loading the root object and its mementos from PM when the application is started;
\item for each root memento, creating a thread to execute a root operation with the memento;
\item should a thread crash, re-creating a new thread that resumes the operation with the same memento; and
\item servicing safe memory reclamation (see \cref{sec:reclamation}
  for details).
\end{enumerate*}
As a caveat, we currently assume
only a single Intel Optane DCPMM module is utilized; and
the type of root object and mementos and the number of root mementos are statically known.
Both of them, while unrealistic, can be easily lifted by designing a more sophisticated monitor, which we leave as future work.

\paragraph{Timestamp Calibration}

At an application start, we need to calibrate timestamps because those retrieved by Intel-x86's \code{rdtsc} and \code{rdtscp} instructions measure the number of ticks after the system was booted, and as a result, they are not strictly increasing in case of reboots either due to crashes or by user commands.
This invalidates our assumption on timestamp monotonicity for \textsc{Checkpoint} and the other primitive operations we will see shortly.
We re-establish this assumption by calibrating timestamps at application starts as follows:
\begin{enumerate*}
\item[\ding{182}] we calculate the maximum timestamp checkpointed in all the mementos, say $T_{\textrm{max}}$;
\item[\ding{183}] we get the current timestamp retrieved by Intel-x86's \code{rdtsc} instruction, say $T_{\textrm{init}}$; and
\item[\ding{184}] we add the offset $(- T_{\textrm{init}} + T_{\textrm{max}})$ to all the timestamps retrieved by \code{rdtsc} and \code{rdtscp} in the application execution.
\end{enumerate*}
Then it is straightforward that the calibrated timestamps are always larger than those checkpointed in the application's past executions.


%% file: cas.tex
\section{Detectable Compare-and-Swap}
\label{sec:cas}

We design, as a key primitive operation, detectable CAS on atomic PM location.
Our algorithm is based on those of Attiya \etal{}~\cite{nrl} and Ben-David \etal{}~\cite{nrlp}.
Their key idea is performing architecture-provided plain CASes twice:
it first performs a plain CAS that annotates the thread id to the location's pointer value,
checkpoints the operation so that it can be detected later, and
performs another plain CAS that removes the annotation from the pointer value.
If a load or CAS operation sees a pointer value annotated with a thread id---that means there is an ongoing CAS, then it \emph{helps} its second plain CAS in a detectable manner.
The detectable helping mechanism requires an array of $O(P^2)$~\cite{nrl} or $O(P)$~\cite{nrlp} sequence numbers for each object in PM, where $P$ is the number of threads, prohibiting its use for space-efficient DSs such as hash tables.
We overcome this limitation by proposing a new helping mechanism that requires only an architecture word for each object and a global array of $O(P)$ timestamps in DRAM and PM.

\subsection{Components and System Assumptions}


We present the components for detectable CAS and their assumptions on the underlying system platform.

\paragraph{Location}

A location is just an architecture word, which is 64 bits in the Intel Xeon platforms supporting DCPMM.
We split 64 bits into three categories: 1-bit \emph{parity} (for helping, see \cref{sec:cas:help} for details), 9-bit thread id, 9-bit user tag, and 45-bit offset.
The thread id 0 is reserved for the CAS algorithm's purposes.
Thread id can distinguish 511 threads and offset can address 32TB, which are sufficient for the current generations of DCPMM.
(We may need to allocate more bits to them by shrinking user tag in the future, though.)
The tag~\cite{tagging} is reserved for users to annotate arbitrary bits to pointer values for algorithm correctness (\eg{}, the synchronization in Harris's list~\cite{harris}) or optimization (\eg{}, hash value in the Clevel hash table~\cite{clevel}).
We assume the \textsc{encode} and \textsc{decode} functions convert a triple of parity, thread id, offset with user tag into a 64-bit word and vice versa.

A location provides two operations: load and detectable CAS.
Both operations are oblivious of parity and thread id so that their input and output pointer values should be annotated with the default values, namely evenness and the reserved thread id 0.
These operations differ from the architecture-provided plain load and CAS operations in that they additionally receive the current thread id as an argument, and CAS also receives a memento.

\paragraph{Memento}

A CAS memento is also a 64-bit architecture word split into three catagories: 1-bit parity, 1-bit failure flag, and 62-bit timestamp retrieved by Intel-x86's \code{rdtscp} instruction.
We reserve timestamp 0, which cannot appear after booted, for those mementos that did not checkpoint any operations yet.
The timestamp is used to calculate $T_{\textrm{max}}$ for timestamp calibration when the application is started (\cref{sec:overview:application}).
The 62 bits are sufficient to support systems running for about 47 years without incurring integer overflow.
We assume the \textsc{encodeTs} and \textsc{decodeTs} functions convert a tuple of parity and timestamp into a 64-bit word and vice versa.

For correct synchronization, we assume \code{rdtscp} guarantees the following properties that are indeed satisfied by the Intel Xeon platforms supporting DCPMM:
\begin{enumerate*}
\item \emph{Synchronous}: the timestamp is synchronized among multiple CPU cores even for multiprocessor systems.
  This property is ensured in modern platforms by synchronous reset signal from the mainboard.
\item \emph{Constant}: the timestamp increments at a constant rate even when CPU frequency changes.
  This property is ensured in modern platforms by dedicated timestamp generators.
  Thanks to this property, \eg{}, two consecutive invocations of \code{rdtscp} return different timestamps.
\item \emph{Serializing}: \code{rdtscp}, when followed by \code{sfence}, is serializing so that the surrounding instructions are never reordered across these instructions.
  This property is described in Intel-x86's architecture manual~\cite{intel-manual}.
  We use this property to ensure the correctness of our helping mechanism (see \cref{sec:cas:help} for more details).
\end{enumerate*}


\paragraph{Global Timestamp Arrays}

We introduce three more global arrays of $P$ timestamps annotated with 1-bit parities: $OWN$ in DRAM and $HELP\textit{[2]}$ in PM.
The former records the timestamp of the last CAS operation performed by each thread, and is initialized as the maximum timestamp checkpointed in CAS mementos at application starts.
The latter checkpoints the maximum timestamp of each thread's CAS operations helped by the other threads for each parity.
As we will see, these invariants are maintained and exploited in our algorithm.
The $HELP\textit{[2]}$ array is initialized with 0 when an application is created and used to calculate $T_{\textrm{max}}$ for timestamp calibration because they are retrieved in post-crash recovery executions.



\subsection{Normal Execution}

\begin{algorithm}[t]
\caption{Atomic Pointer Location supporting Detectable CAS}\label{alg:cas:base}
\begin{algorithmic}[1]
  \Function{Load}{$loc,tid$}
  \State{$cur \gets \Call{LoadPlain}{loc}$}
  \State $\textbf{return}~\Call{LoadHelp}{loc,cur}$
  \EndFunction
  \Function{CAS}{$loc,old,new,tid,mmt,recovery$}
  \State $(p_{own}, t_{own}) \gets \Call{decodeTs}{\textsc{LoadPlain}(OWN[tid])}$
  \State $old' \gets \Call{encode}{\textsc{Even},0,old}$
  \State $new' \gets \Call{encode}{\neg p_{own},tid,new}$
  \If{$recovery$}
  \State{$pt \gets \Call{LoadPlain}{mmt}$}
  \State{$(p_{mmt},t_{mmt}) \gets \textsc{decodeTs}(pt)$}
  \State{$cur \gets \Call{LoadPlain}{loc}$}
  \State $(\_, t_{local}) \gets \Call{decodeTs}{\textsc{LoadPlain}(LOCAL[tid])}$
  \IfThen{$t_{mmt} < t_{local}$}{$\textbf{goto}~21$}
  \If{$p_{mmt} = \textsc{Fail}$}
  \State{$\Call{StorePlain}{LOCAL[tid],pt}$}
  \State{$\textbf{return}~(\textsc{Err}~\Call{LoadHelp}{loc,cur})$}
  \EndIf
  \IfThen{$t_{mmt} \neq 0 \land t_{mmt} < t_{own}$}{$ts_{succ} \gets pt; \textbf{goto}~39$}
  \IfThen{$t_{mmt} \neq 0 \land t_{mmt} \ge t_{own}$}{$ts_{succ} \gets pt; \textbf{goto}~36$}
  \State{$(p_{cur}, tid_{cur}, o_{cur}) \gets \Call{decode}{cur}$}
  \IfThen{$tid_{cur} = tid \land o_{cur} = new$}{$\textbf{goto}~33$}
  \State $t_{help} \gets \Call{LoadPlain}{HELP[\neg p_{own}][tid]}$
  \IfThen{$t_{own} < t_{help}$}{$\textbf{goto}~34$}
  \EndIf
  \If{$\textsc{CASPlain}(loc,old',new') ~\textbf{is}~ (\textsc{Err}~cur)$}
  \State $cur \gets \Call{LoadHelp}{loc,cur}$
  \IfThen{$cur = old$}{$\textbf{goto}~26$}
  \State{$ts_{fail} \gets \textsc{encodeTs}(\textsc{Fail},0)$}
  \State{$\Call{StorePlain}{mmt,ts_{fail}}; \textbf{flush}~mmt$}
  \State{$\Call{StorePlain}{LOCAL[tid],ts_{fail}}; \textbf{return}~(\textsc{Err}~cur)$}
  \EndIf
  \State{$\textbf{flush}~loc$}
  \State{$ts_{succ} \gets \Call{encodeTs}{\neg p_{own}, \textbf{rdtscp}}; \textbf{lfence}$}
  \State{$\Call{StorePlain}{mmt,ts_{succ}}; \textbf{flushopt}~mmt$}
  \State{$\Call{StorePlain}{OWN[tid],ts_{succ}}$}
  \State{$new'' \gets \Call{encode}{\textsc{Even},0,new}$}
  \IfThen{$\Call{CASPlain}{loc, new', new''}$ ~\textbf{is}~ \textsc{Err}}{$\textbf{sfence}$}

  \State{$\Call{StorePlain}{LOCAL[tid],ts_{succ}}; \textbf{return}~\textsc{Ok}$}
  \EndFunction
  \Function{CASMementoClear}{$mmt$}
  \State{$\Call{StorePlain}{mmt,\textsc{encodeTs}(\textsc{Even},0)}$}
  \State{$\textbf{flushopt}~mmt$}
  \EndFunction
\end{algorithmic}
\end{algorithm}

\cref{alg:cas:base} presents the load and detectable CAS algorithms.
We omit memory orderings, but they are appropriately annotated in our implementation.
The \textsc{Load} operation (L1) simply performs an architecture-provided plain load.
Since it may observe a transient value written by a concurrent (detectable) CAS operation's first plain CAS (L26),
it invokes \textsc{LoadHelp} to help the CAS operation and retrieve a stable value without parity or thread id annotations (see \cref{sec:cas:help} for more details).
On the other hand, the \textsc{CAS} operation (L5) first checks if it is executed in the recovery mode (see \cref{sec:cas:recovery} for more details).
Otherwise, \textsc{CAS}
\begin{enumerate*}
\item[\ding{182}] reads the parity and timestamp of its last CAS operation (L6) and toggles the parity ($\neg p_{own}$ at L8 and L34);
\item[\ding{183}] performs the first plain CAS from the old pointer (L7) to the new one (L8) annotated with the toggled parity and its thread id (L26);
\item[\ding{184}] if the plain CAS fails, then retrieves a stable current value using \textsc{LoadHelp}, and depending on the current value, retries the plain CAS, or invalidates the memento by the sentinel value $\textsc{encodeTs}(\textsc{Fail},0)$ and returns (L27-L31);
\item[\ding{185}] flushes the first plain CAS (L33);
\item[\ding{186}] checkpoints the toggled parity and the current timestamp in the memento (L34-L35); 
\item[\ding{187}] maintains $OWN$'s invariant (L36); and
\item[\ding{188}] tries to perform the second plain CAS removing parity and thread id annotations, and regardless of its result, ensures the checkpointing is flushed (L38). 
  The second plain CAS may fail because a concurrent thread may have already helped performing it.
\end{enumerate*}
%
%
A memento can later be reused after clear to the reserved timestamp 0 in a persistent manner (L41).

In the absence of crash and contention with the other threads, CAS effectively replaces an old pointer value with the new one (without parity or thread id annotations) in a persistent manner.
Now we will see what would happen in case of crash (\cref{sec:cas:recovery}) and contention (\cref{sec:cas:help}).


\subsection{Recovery}
\label{sec:cas:recovery}

Should a thread crash in the middle of a CAS, the post-crash recovery must resume the operation so that CAS is performed exactly once.
In particular, the recovery execution of CAS needs to distinguish the cases when the pre-crash execution is interrupted
\begin{enumerate*}
\item before persisting the first plain CAS (L26);
\item after an unsuccessful first plain CAS (L30); or
\item after a successful first plain CAS and that and before persisting the memento (L35); or
\item after that.
\end{enumerate*}
To this end, the recovery execution of CAS
\begin{enumerate*}
\item[\ding{182}] globally initializes $OWN$ only once for each thread as the maximum checkpointed timestamp of the thread's CAS mementos;
\item[\ding{183}] loads and decodes the memento (L10-L11);
\item[\ding{184}] if the memento's checkpointed timestamp is less than the thread's $LOCAL$, then go to step \ding{188} (L12-L14); 
\item[\ding{185}] if the memento checkpointed a failure, then reports an error (L15-L18);
\item[\ding{186}] if the memento checkpointed a CAS operation that is not the thread's last one, then return \textsc{Ok} (L19);
\item[\ding{187}] if the memento checkpointed the thread's last CAS operation, then resumes from the normal execution's step \ding{187} (L20);
\item[\ding{188}] if the location contains the first CAS's new value, then resumes from the normal execution's step \ding{185} (L22);
\item[\ding{189}] if the CAS is helped by a concurrent thread (see \cref{sec:cas:help} for more details), then resumes from the normal execution's step \ding{186} (L24); and finally,
\item[\ding{190}] resumes the normal execution (L25). 
\end{enumerate*}

For each case, the post-crash recovery execution correctly resumes the operation as follows:
\begin{enumerate}
\item Since the operation performed nothing to the location and memento, it goes to the step \ding{184} and resumes the normal execution.
%
\item Since the operation checkpoints a failure, it goes to the step \ding{185} and returns \textsc{Err} again.
\item Since the operation performed nothing to the memento, it goes to the step \ding{188}.
  Since the first plain CAS is persisted, either the location still contains the plain CAS's new value (\ding{188}) or the CAS is helped by a concurrent thread (\ding{189}).
  In either case, the recovery execution correctly resumes from the normal execution's step \ding{185} (or \ding{186}, respectively) checkpointing the memento.
\item If the CAS is not the last one, there is nothing to do (\ding{186});
  otherwise, the recovery execution correctly resumes from the normal execution's step \ding{188} performing the second plain CAS.
\end{enumerate}

\subsection{Helping}
\label{sec:cas:help}

\begin{algorithm}[t]
\caption{Atomic Pointer Location supporting Detectable CAS}\label{alg:cas:help}
\begin{algorithmic}[1]
  \Function{LoadHelp}{$loc,old$}
  \State{$(p_{old}, tid_{old}, o_{old}) \gets \Call{decode}{old}$}
  \IfThen{$tid_{old} = 0$}{$\textbf{return}~o_{old}$}
  \State $t_{cur} \gets \textbf{rdtscp}; \textbf{lfence}$
  \State{$cur \gets \Call{LoadPlain}{loc}$}
  \State{$(\_, tid_{cur}, o_{cur}) \gets \Call{decode}{cur}$}
  \IfThen{$tid_{cur} = 0$}{$\textbf{return}~o_{cur}$}
  \IfThen{$old \neq cur$}{$old \gets cur; \textbf{goto}~4$}
  \IfThen{$\textbf{rdtscp} < t_{cur} + PATIENCE$}{$\textbf{goto}~6$}
  \State $t_{help} \gets \Call{LoadPlain}{HELP[p_{old}][tid_{old}]}$
  \IfThen{$t_{cur} \le t_{help}$}{$old \gets \Call{LoadPlain}{loc}; \textbf{goto}~2$}
  \State $\textbf{flushopt}~loc$
  \If {$\Call{CASPlain}{HELP[p_{old}][tid_{old}],t_{help},t_{cur}}$ ~\textbf{is}~ \textsc{Err}}
  \State{$old \gets \Call{LoadPlain}{loc}; \textbf{goto}~2$}
  \EndIf
  \State $\textbf{flushopt}~HELP[p_{old}][tid_{old}]$
  \State $old' \gets \Call{encode}{\textsc{Even},0,o_{old}}$
  \If {$\textsc{CASPlain}(loc,old,old')$ ~\textbf{is}~ (\textsc{Err}~cur)}
  \State $old \gets cur; \textbf{goto}~2$
  \EndIf
  \State $\textbf{return}~old'$
  \EndFunction
\end{algorithmic}
\end{algorithm}

A thread may help an ongoing CAS operation's second plain CAS.
We synchronize a CAS operation and a helper in a detectable manner using timestamp and parity.
The \textsc{LoadHelp} operation, presented in \cref{alg:cas:help}, receives a location $loc$ and an $old$ value of the location, and it
\begin{enumerate*}
\item[\ding{182}] returns $old$ if it is stable (L3);
\item[\ding{183}] reads the current timestamp and executes a fence (L4);
\item[\ding{184}] loads a value, say $cur$, from the location again, and returns it if it is stable (L5-L7);
\item[\ding{185}] if $old \neq cur$, then retries from L4 (L8);
\item[\ding{186}] backs off for a certain period of time, achieving better performance by allowing more chances for the thread performing the ongoing CAS to finish its own operation (L9);
\item[\ding{187}] from the $HELP$ entry with $old$'s parity and thread id, loads the maximum timestamp of the thread's CAS operations helped by the other threads for the parity; and if it is more recent than my timestamp, then retries from the beginning (L10-L11);
\item[\ding{188}] flushes the location (L12);
\item[\ding{189}] tries to increment the $HELP$ entry to my timestamp in a persistent manner; and if it fails, then retries from the beginning (L13-L16);
\item[\ding{190}] helps the ongoing CAS operation's second plain CAS, and if it fails, then retries from the beginning (L17-L20); and finally,
\item[\ding{191}] returns the newly written stable pointer value (L21).
\end{enumerate*}

The $HELP$ array is used to synchronize the \textsc{LoadHelp} and \textsc{CAS} operations in case of crashes.
A post-crash recovery execution of the \textsc{CAS} operation consults the $HELP$ array (L23-L24) to distinguish the cases \ding{172} and \ding{174} in \cref{sec:cas:recovery}, or in other words, whether the pre-crash execution's first plain CAS (L26) is persisted.
To this end, we use the following invariant: suppose a thread $tid$'s $n$-th \textsc{CAS} invocation's parity and timestamp are $p_n$ and $t_n$, respectively.
Obviously, the sequence $\{P_i\}$ is alternating between \textsc{Even} and \textsc{Odd}, and the sequence $\{t_i\}$ is strictly increasing.
\emph{Then $t_{n-1} < HELP[p_n][tid]$ if and only if the $n$-th or a later \textsc{CAS} with the parity $p_n$ is helped by a concurrent thread.}
On the one hand, \textsc{CAS} exploits this invariant to make a correct decision at L23-L24.
On the other hand, \textsc{LoadHelp} maintains this invariant as follows.

\newif\ifdrawbox
\tikzset{
  draw box/.is if flag=drawbox, draw box,
  /litmus/every litmus/.style={
    AArch64,
    /tikz/draw box/.if true=then {
      execute at end scope={
        \draw[dashed, thin, opacity=0.2]
          (current bounding box.south west) rectangle (current bounding box.north east);
      }
    },
    every relation/.append style=overlay,
  },
}

\newcommand{\hblab}{\ensuremath{\textcolor{blue}{hb}}}
\newcommand{\polab}{\ensuremath{\textcolor{black}{po}}}
\newcommand{\rflab}{\ensuremath{\textcolor{red}{rf}}}
\newcommand{\frlab}{\ensuremath{\textcolor{orange}{fr}}}
\newcommand{\hbto}{\ensuremath{\stackrel{\hblab}{\rightarrow}}}
\newcommand{\poto}{\ensuremath{\stackrel{\polab}{\rightarrow}}}
\newcommand{\rfto}{\ensuremath{\stackrel{\rflab}{\rightarrow}}}
\newcommand{\frto}{\ensuremath{\stackrel{\frlab; \rflab^?}{\rightarrow}}}

\begin{figure}
  \centering
  \tikzset{draw box=false}
  \lstset{language=AArch64, style=litmus, basicstyle=\footnotesize\ttfamily}%
  \begin{tikzpicture}[/litmus/append to tikz path search,kind=full]
    \node[instructions=0] (cas) {
      !\node { }; \\
      \node[event=a] {\:t\textsubscript{n-1}=rdtscp;\,lfence}; \\
      \node[event=b] {\:Update\textsubscript{n-1,2}\,/\,Load}; \\[0.3em]
      \node[event=c] {\:Update\textsubscript{n,1}\,}; \\
      \node[event=d] {\:t\textsubscript{n}}; \\
      \node[event=e] {\:Update\textsubscript{n,2}\,/\,Load\,}; \\[0.3em]
      \node[event=f] {\:Update\textsubscript{n+1,1}}; \\
      \node[event=g] {\:t\textsubscript{n+1}}; \\
    };
    \draw (a) edge[po, out=-180, in=180, swap] (c);
    \draw (e) edge[po, out=-180, in=180, swap] (g);
    \node[thread header, above=of cas, below delimiter=|, inner sep=0.05cm] {CAS};

    \node[instructions=1, align=left] (loadhelp) {
      !\node { }; \\
      !\node { }; \\
      !\node { }; \\[0.3em]
      \node[event=h, inner sep=1pt] {\:Load}; \\
      \node[event=i, inner sep=1pt] {\:t\textsubscript{h}=rdtscp;\,lfence}; \\
      \node[event=j, inner sep=1pt] {\:Load}; \\[0.3em]
    };
    \draw (h) edge[po, out=0, in=90, label pos=0.5] (i);
    \draw (i) edge[po, out=-90, in=0, label pos=0.5] (j);
    \node[thread header, above=of loadhelp, below delimiter=|, inner sep=0.05cm] {LoadHelp};
  
    \draw[yshift=0.58cm, xshift=-1.3cm, thick] (0,0) -- (2.7,0);
    \draw[yshift=-0.73cm, xshift=-1.3cm, thick] (0,0) -- (2.7,0);
    \begin{scope}[execute at begin node=$, execute at end node=$]
      !\node[draw, align=left, yshift=-1.4cm, xshift=3.5cm, inner sep=0.1cm] {
        \textcolor{darkgray}{a} \,\hbto\, \textcolor{darkgray}{c} \,\hbto\, \textcolor{darkgray}{h} \,\hbto\, \textcolor{darkgray}{i} \,\hbto\, \textcolor{darkgray}{j} \,\hbto\, \textcolor{darkgray}{e} \,\hbto\, \textcolor{darkgray}{g} };
    \end{scope}
  
    \draw (c) edge[rf, label text={\ensuremath{\textcolor{red}{rf}}}] (h);
    \draw (c) edge[rf, label text={\ensuremath{\textcolor{red}{rf}}}, out=-10, in=170] (j);
    \draw (j) edge[fr, label text={\ensuremath{\textcolor{orange}{fr}}\textcolor{black}{;}\ensuremath{\textcolor{red}{rf}\textcolor{black}{\textsuperscript{?}}}}] (e);
  \end{tikzpicture}

  \caption{Synchronization of Detectable CAS and LoadHelp}
  \label{fig:cas:synchronization}
\end{figure}
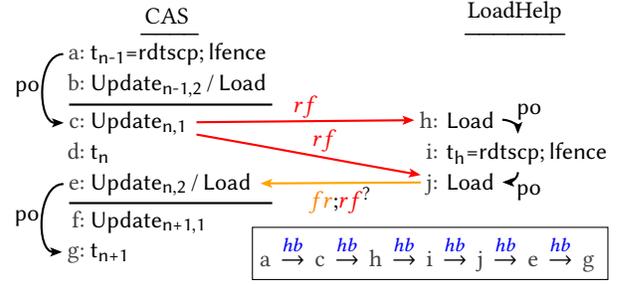




Suppose a \textsc{LoadHelp} operation retrieves a timestamp $t_h$ at L4 and tries to help a the second plain CAS of a thread $tid$'s $n$-th \textsc{CAS} invocation, as illustrated in \cref{fig:cas:synchronization}.
Here, we depict the plain CASes and timestamp retrieval of $tid$'s $(n-1)$-th to $(n+1)$-th \textsc{CAS} invocations and loads and timestamp retrieval of a\textsc{LoadHelp} invocation,
where $\textsl{Update}_{n,i}$ represents the $i$-th plain CAS of $tid$'s $n$-th CAS and $\textsl{Load}$ represents a load from a location.
Then we have the following properties:
\begin{enumerate*}
\item $t_{n-1} < t_h$ from $a \poto c \rfto h \poto i$; and
\item $t_h < t_{n+1}$ from $i \poto j \frto e \poto g$,
\end{enumerate*}
where $\polab$ is the program order; $\rflab$ is the read-from relation; $\frlab$ is the from-read relation; $\frlab; \rflab^?$ is the from-read relation possibly followed by a read-from relation; and all of them constitute the \emph{happens-before} relation $\hblab$ in the x86-TSO memory model~\cite{x86-tso-tphol}.
For more details on the x86-TSO memory model, we refer the readers to \cite{x86-tso-tphol} for space purposes.

Now recall that \textsc{LoadHelp} persists $\textsl{Update}_{n,1}$ (L12), atomically increases $HELP[p_n][tid]$ to $t_h$ (L13-L16), and helps $U_{n,2}$ (L17-L18).
If the thread $tid$'s $n$-th CAS is helped by a concurrent thread, then we have $t_{n-1} < t_h \le HELP[p_n][tid]$ due to the property \textbf{(1)};
conversely, if $t_{n-1} < HELP[p_n][tid]$, then it should be the result of a help neither
\begin{enumerate*}
\item for $tid$'s CASes with the parity $\neg p_n$; nor
\item for $tid$'s $(n-2)$-th or earlier CASes due to the property \textbf{(2)}.
\end{enumerate*}
In other words, \textsc{LoadHelp} maintains the invariant of $HELP$ mentioned above.



%% file: smo.tex
\section{Detectable Insertion and Deletion}
\label{sec:smo}

We design, as optimized primitive operations, detectable insertion and deletion of nodes on atomic PM location.
While CAS is a general primitive operation for pointer-based DSs, we observe that the performance of DSs implemented with detectable CAS~(\cref{sec:cas}) are sometimes significantly worse than that of hand-tuned detectable DSs.
The primary reason is that the general detectable CAS performs plain CASes to the same location twice and flushes the location between them, incurring an extremely high contention on the location among multiple threads.

As an optimization, we design a more efficient atomic pointer location object supporting detectable \emph{insert}---\emph{atomically replacing the \textsc{Null} pointer}---and \emph{delete}---\emph{atomically detaching a valid memory block from DSs}---operations by capturing the essence of hand-tuned detectable DSs.
The key idea is distributing the contention of multiple threads into multiple memory blocks, significantly relieving contention on any single location.
Such an optimization, however, requires non-trivial synchronization among multiple memory blocks,
thus limiting its application to only insert and delete operations.
While less general than CAS, insert and delete operations still support a wide variety of DSs, \eg{}, we can implement Michael-Scott's queue (MSQ)~\cite{msq} and exchanger with insert and delete operations.



\subsection{Components and DS Assumptions}


We present the components for atomic pointer location supporting insertion and deletion and their assumptions on the DS.

\paragraph{Location}

A location is a 64-bit architecture word, which we split into four categories: 1-bit ``persist'' flag for the \emph{link-and-persist} technique~\cite{link-persist}, 8 bits reserved for future purposes, 10-bit user tag, and 45-bit offset.
We follow David~\etal{}~\cite{link-persist} in enforcing the invariant that, if the persist bit is cleared, then the pointer value is persisted.
Such an invariant is cooperatively maintained by location's three operations: load, insertion, and deletion.
The load operation ensures that the returned pointer value is always persisted in the location.
Insertion receives a location and a new pointer value as arguments, and deletion receives a location, old and new pointer values, a memento, and the current thread id.
We assume the \textsc{encode} and \textsc{decode} functions convert a tuple of persist bit and offset with user tag into an 64-bit word and the other way around.

\paragraph{Memory Block}

We assume each memory block has a dedicated 64-bit architecture word, which we call \code{repl}, that describes the memory block that replaces itself as an atomic location's next pointer value.
We split 64 bits into two categories: 9-bit thread id, 10-bit user tag, and 45-bit offset.
Similarly to \cref{sec:cas}, the thread id 0 is reserved for those blocks that are not replaced yet.
We assume the \textsc{encodeR} and \textsc{decodeR} functions convert a tuple of thread id and offset with user tag into an 64-bit word and the other way around.

\paragraph{Traversal}

As we will see, we checkpoint the insertion of a memory block to a DS in the inserted node's \code{repl} field, and
for efficiency purposes, we perform such checkpoint only when the block is deleted~(see \cref{sec:smo:normal} for details).
Should the system crash, we detect whether a non-checkpointed memory block was inserted before the crash by checking if the block is still in the DS.
To this end, we require the ability to traverse all the memory blocks in a DS.
For instance, an MSQ~\cite{msq} is traversable from its head by recursively chasing each node's next pointer.

\paragraph{Memento}

Insertion and deletion do not need a memento because the progress is checkpointed in the input node's \code{repl} field.

\subsection{Normal Execution}
\label{sec:smo:normal}

\begin{algorithm}[t]
\caption{Load, Insertion, and Deletion}\label{alg:smo:base}
\begin{algorithmic}[1]
  \Function{Load}{$loc$}
  \State{$old \gets \Call{LoadPlain}{loc}$}
  \State{$(p_{old}, o_{old}) \gets \Call{decode}{old}$}
  \IfThen{$\neg p_{old}$}{$\textbf{return}~o_{old}$}
  \State{$t \gets \textbf{rdtsc}$}
  \State{$cur \gets \Call{LoadPlain}{loc}$}
  \State{$(p_{cur}, o_{cur}) \gets \Call{decode}{cur}$}
  \IfThen{$\neg p_{cur}$}{$\textbf{return}~o_{cur}$}
  \IfThen{$old \neq cur$}{$old \gets cur; \textbf{goto}~3$}
  \IfThen{$\textbf{rdtsc} < t + PATIENCE$}{$\textbf{goto}~6$}
  \State{$\textbf{flush}~loc$}
  \State{$old' \gets \Call{encode}{\textsc{False}, o_{old}}$}
  \IfThen{$\Call{CASPlain}{loc,old,old'} ~\textbf{is}~ (\textsc{Err}~cur)$}{$old \gets cur; \textbf{goto}~3$}
  \State{$\textbf{return}~old'$}
  \EndFunction
  \Function{Insert}{$loc,new,ds,recovery$}
  \If{$recovery$}
  \IfThen{$\Call{Contains}{ds,new} \lor (\Call{LoadPlain}{new.\code{repl}} \neq \textsc{Null})$}
  {$\textbf{return}~\textsc{Ok}$}
  \State{$~\textbf{return}~(\textsc{Err}~\Call{Load}{loc})$}
  \EndIf
  \If{$\Call{CASPlain}{loc,\textsc{Null},\textsc{encode}(\textsc{True},new)} ~\textbf{is}~ \textsc{Err}$}
  \State{$\textbf{return}~(\textsc{Err}~\Call{Load}{loc})$}
  \EndIf
  \State{$\textbf{flush}~loc$}
  \State{$\Call{CASPlain}{loc,\textsc{encode}(\textsc{True},new),\textsc{encode}(\textsc{False},new)}$}
  \EndFunction
  \Function{Delete}{$loc,old,new,tid,recovery$}
  \If{$recovery$}
  \State{$(tid_{new}, o_{new}) \gets \textsc{decodeR}(\Call{LoadPlain}{old.\code{repl}})$}
  \IfThen{$tid \neq tid_{new}$}{$\textbf{return}~(\textsc{Err}~\Call{LoadHelp}{loc,old})$}
  \State{$\textbf{goto}~37$}
  \EndIf
  \State{$new' \gets \textsc{encodeR}(tid, new)$}
  \If{$\Call{CASPlain}{old.\code{repl}, \textsc{Null}, new'} ~\textbf{is}~ (\textsc{Err}~cur)$}
  \State{$\textbf{return}~(\textsc{Err}~\Call{LoadHelp}{loc,old})$}
  \EndIf
  \State{$\textbf{flushopt}~old.\code{repl}$}
  \State{$\Call{CASPlain}{loc,old,new}; \Call{DeferFlush}{loc}; \Call{Retire}{old}$}
  \State{$\textbf{return}~old$}
  \EndFunction
\end{algorithmic}
\end{algorithm}

\cref{alg:smo:base} presents the load, insertion, and deletion algorithms.

\paragraph{Load}

The \textsc{Load} operation, which we adopt from \cite{link-persist}, ensures the returned pointer values are always persisted by
\begin{enumerate*}
\item[\ding{182}] performing an architecture-provided plain load and decodes the pointer value (L2-L3);
\item[\ding{183}] if its persist bit is cleared---which implies the pointer value is persisted, then returning the pointer value (L4); otherwise,
\item[\ding{184}] retrying for a while to read a pointer value without the persist bit being set (L5-L10); and if it fails,
\item[\ding{185}] flushing $loc$ itself (L11);
\item[\ding{186}] trying to clear the persist bit of $loc$ by performing a CAS, and if it fails, retrying from the beginning (L12-L13); and
\item[\ding{187}] returning the pointer value with the persist bit being cleared (L14).
\end{enumerate*}

\paragraph{Insertion}

The \textsc{Insert} operation on a location, say $loc$, atomically replaces the \textsc{Null} pointer with the given pointer value, say $new$, in a persistent manner (L16).
\textsc{Insert} first checks if it is executed in the recovery mode (see \cref{sec:smo:recovery} for more details); otherwise, it
\begin{enumerate*}
\item[\ding{182}] atomically updates $loc$'s pointer value from the \textsc{Null} pointer to the given pointer with the persist bit being set (L21); if it fails, reports an error with the current pointer value (L22);
\item[\ding{183}] flushes $loc$ (L24); and
\item[\ding{184}] tries to atomically clear $loc$'s persist bit (L25).
\end{enumerate*}
It is okay to let the second CAS fail (L25) because it means a concurrent load or delete operation should have persisted $loc$ and cleared the persist bit.

\paragraph{Deletion}

\begin{figure}
  \centering
  \includesvg[width=\columnwidth]{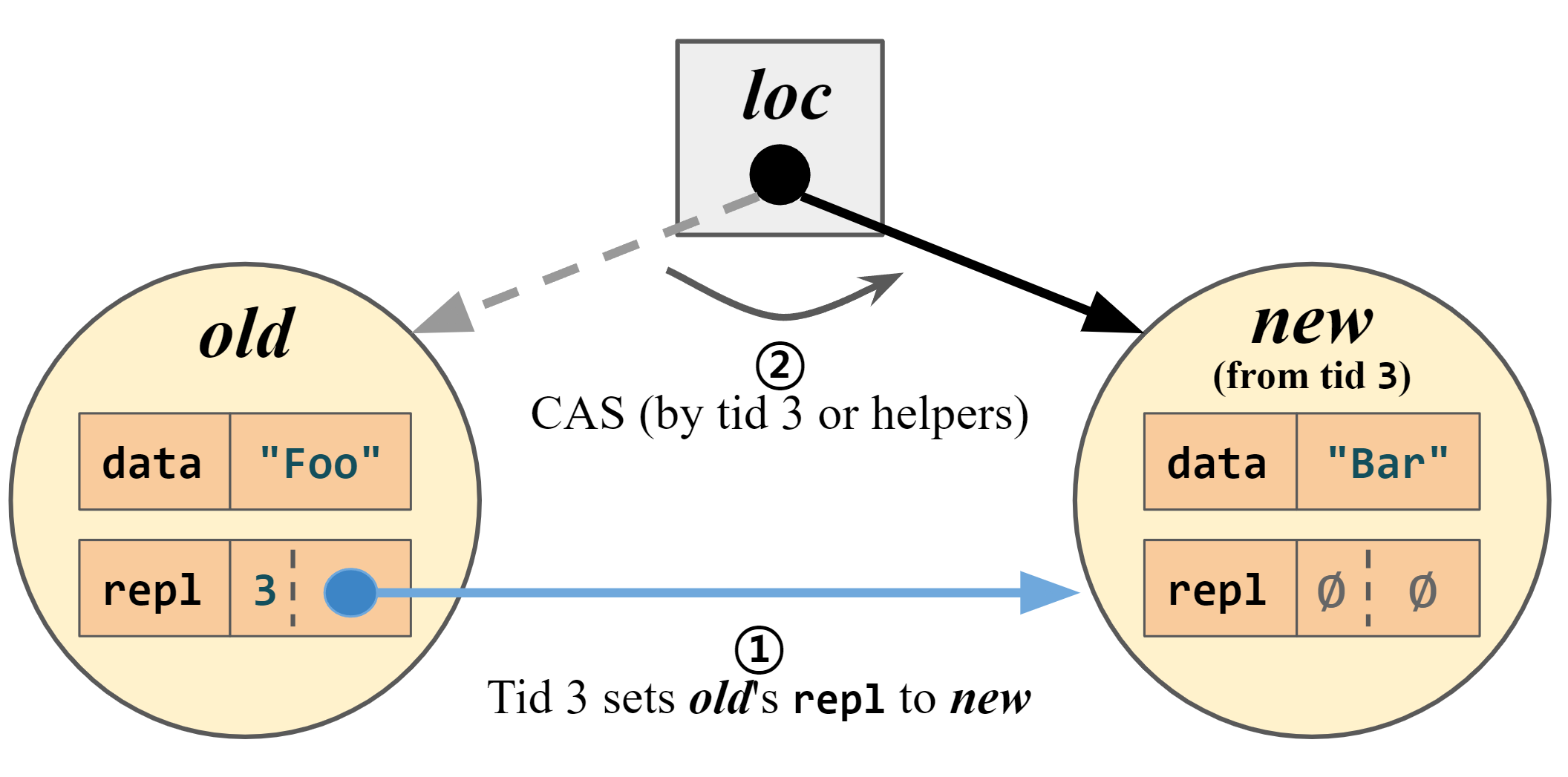}
  \caption{Delete Operation Procedure}
  \label{fig:smo:delete}
  \vspace{-0.7em}
\end{figure}

The \textsc{Delete} operation, illustrated in \cref{fig:smo:delete}, on a location, say $loc$, atomically detaches the given pointer to a valid memory block, say $old$, and stores another given pointer, say $new$, in a persistent manner (L27).
\textsc{Delete} first checks if it is executed in the recovery mode; otherwise, it
\begin{enumerate*}
\item[\ding{182}] tries to atomically install $new$ annotated with the current thread id to $old$'s \code{repl} field (L33-L34); if it fails, helps the completion of concurrent delete operations and reports an error (L35, see \cref{sec:smo:help} for details on helping);
\item[\ding{183}] flushes $old$'s \code{repl} field (L37);
\item[\ding{184}] tries to replace $loc$'s value from $old$ to $new$, persists $loc$ in a \emph{deferred} manner, and \emph{retires} $old$ (L38); and
\item[\ding{185}] returns $old$ (L39).
\end{enumerate*}
Here, we retire $old$ so that it will be freed once it is no longer accessible from the other threads using safe memory reclamation schemes such as hazard pointers~\cite{hp} and epoch-based reclamation~\cite{ebr}.
We also ensure $loc$ is flushed at least before $old$ is freed using \textsc{DeferFlush} so that $loc$ points to a valid memory block even in case of crashes (see \cref{sec:reclamation}
for details on safe reclamation).

A delete operation is committed when the CAS on $old$'s \code{repl} (L34) is persisted, while its effects are applied to $loc$ later (L38).
It is still safe for concurrent operations to see an old value of $loc$ even though a new value is already committed:
a concurrent load operation can simply return the old value because it can linearize before the deletion;
similarly, a concurrent insert and delete operations can linearize beforehand and fail.


\subsection{Recovery}
\label{sec:smo:recovery}

\paragraph{Insertion}

The recovery execution of insertion needs to distinguish the cases when the pre-crash execution is interrupted before or after persisting the first CAS (L21).
%
%
To this end, we apply \emph{direct tracking} method in \cite{tracking}, \ie{},
the recovery execution of insertion checks if either
\begin{enumerate*}
\item the node is still contained in the enclosing DS by performing a traversal; or
\item the node's \code{repl} field is populated, which means it is already deleted (L18).
\end{enumerate*}
It is straightforward that it is the case if and only if the node have indeed been inserted.
Here, the recovery execution spuriously fails to insert the given memory block---even though $loc$ is \textsc{Null}---when the pre-crash execution is interrupted before persisting the first CAS.
In this case, the memento needs to restart the insert operation in a normal execution.

\paragraph{Deletion}

The recovery execution of deletion needs to distinguish the cases when the pre-crash execution is interrupted
\begin{enumerate*}
\item before persisting the first CAS (L34); or
\item after that.
\end{enumerate*}
To this end, the recovery execution of deletion
\begin{enumerate*}
\item[\ding{182}] loads and decodes the $old$'s \code{repl}, and if its $tid$ is not the current thread id, reports an error (L29-L30); otherwise,
\item[\ding{183}] resumes from the normal execution's step \ding{183} (L31).
\end{enumerate*}

For each case, the post-crash recovery execution correctly resumes the operation as follows:
\begin{enumerate}
\item Since the operation performed nothing to $old$'s \code{repl}, it goes to L30 and spuriously fails the operation;
\item Since $new$ should contain the current thread id, it goes to L37 and correctly resumes from the normal execution's step \ding{183}.
\end{enumerate}

\subsection{Helping}
\label{sec:smo:help}

\begin{algorithm}[t]
\caption{Helping Load}\label{alg:smo:help}
\begin{algorithmic}[1]
  \Function{LoadHelp}{$loc,old$}
  \IfThen{$old = \textsc{Null}$}{$\textbf{return}~(\textsc{Ok}~old)$}
  \State{$new \gets \Call{LoadPlain}{old.\code{repl}}$}
  \State{$(tid_{new}, o_{new}) \gets \Call{decodeR}{old}$}
  \IfThen{$o_{new} = \textsc{Null}$}{$\textbf{return}~(\textsc{Ok}~old)$}
  \State{$t \gets \textbf{rdtsc}$}
  \State{$cur \gets \Call{LoadPlain}{loc}$}
  \IfThen{$cur \ne old$}{$\textbf{return}~(\textsc{Ok}~cur)$}
  \IfThen{$\textbf{rdtsc} < t + PATIENCE$}{$\textbf{goto}~6$}
  \State{$\textbf{flushopt}~old.\code{repl}$}
  \If{$\Call{CAS}{loc, old, o_{new}} ~\textbf{is}~ (\textsc{Err}~e)$}
  {$\textbf{return}~(\textsc{Ok}~e)$}
  \Else
  {$~\textbf{return}~(\textsc{Ok}~o_{new})$}
  \EndIf
  \EndFunction
\end{algorithmic}
\end{algorithm}

A delete operation may help an ongoing concurrent delete operation's second CAS.
Such a help is essential for lock freedom because the concurrent operation may be committed---have successfully performed the first CAS---while its effects have not been applied to $loc$ yet.
%
The \textsc{LoadHelp} operation, presented in \cref{alg:smo:help}, receives a location $loc$ and an old value $old$ of the location and returns a stable value of $loc$ by possibly helping the second CAS of ongoing deletions as follows: it
\begin{enumerate*}
\item[\ding{182}] returns $old$ if it is the stable value \textsc{Null} (L2);
\item[\ding{183}] loads and decodes $old.\code{repl}$ as $new$ (L3-L4);
\item[\ding{184}] returns $old$ if \code{repl} is \textsc{Null} and thus $old$ is stable (L5);
\item[\ding{185}] tries to read a pointer value for a while (L6-L9);
\item[\ding{186}] flushes $old.\code{repl}$ (L10);
\item[\ding{187}] tries to update $loc$ from $old$ to $new$, and regardless of the result, returns the current value of $loc$ (L11-L13).
\end{enumerate*}
%




%% file: volatile.tex
\section{DRAM Scratchpad Optimization}
\label{sec:volatile}

We propose a guideline for optimizing detectable DSs with DRAM scratchpad while preserving detectability.
By DRAM scratchpad we mean putting a part of persistent DSs in DRAM.
Such an optimization is pioneered in SOFT~\cite{soft} and Mirror~\cite{mirror} that replicate PM contents in DRAM for higher performance, and is known to be extremely effective for lock-free DSs in PM.
However, the optimization is neither supported by our composition (\cref{sec:overview}) nor explicitly recognized and exploited for detectable DSs in the literature.
We explain the concept of DRAM scratchpad optimization and propose a correctness guideline for detectable DSs using Michael-Scott's queue (MSQ)~\cite{msq} and SOFT hash table~\cite{soft} as running examples.






\paragraph{Michael-Scott's Queue}

A volatile MSQ~\cite{msq} is a linked list and has ``tail'' pointer that points to the list's tail block in a best-effort basis.
If a volatile MSQ is converted to a detectable MSQ with our transformation~(\cref{sec:overview:transformation}), then the tail pointer becomes a PM location and all its modifications become detectable CASes, incurring noticeable overhead over volatile MSQ.
But the prior works on persistent MSQs~\cite{friedman,dss} made an observation that the tail pointer does not need to be persisted in PM as its exact value is not crucial and it only needs to maintain the following invariants:
the tail is reachable from the queue's ``head'' pointer (also an invariant of the volatile MSQ); and
all the links in the traversal from the head to the tail is persisted (new in the persistent MSQ).
As a result, they put the tail in DRAM scratchpad and ensure the invariants are maintained throughout the execution of operations by
\begin{enumerate*}
\item assigning its head to its tail at application starts;
\item updating its head to its next block only if the tail is different from the head; and
\item updating its tail to its next block only if the link to the next block is persisted.
\end{enumerate*}

Since the invariants already ensure the links from the head to the tail are persisted, load operations do not need to ensure the same property themselves, reducing a flush instruction for each load.
We observe that this also improves the performance significantly.

\paragraph{Guideline}

More generally, a detectable DS with DRAM scratchpad should be accompanied by relational invariants on DRAM scratchpad and PM data, and the invariants should be
\begin{enumerate*}
\item established at application starts;
\item preserved at primitive detectable operations; and
\item preserved at DRAM scratchpad updates.
\end{enumerate*}
We observe that the condition \textbf{(3)} is subtle for recovery executions where stale values checkpointed in a pre-crash execution may be used to update DRAM scratchpad, breaking its invariants.
We should avoid such a case by updating DRAM scratchpad only with latest information, \eg{}, we update MSQ's tail only in normal executions.

\paragraph{SOFT}

Roughly speaking, SOFT~\cite{soft} can be understood as a transformation of two copies of hash tables of which one is moved to PM and the other remains in DRAM as a scratchpad.
The invariant is the equivalence of the two copies, and SOFT ensures this invariant by
\begin{enumerate*}
\item reconstructing a DRAM copy from the PM copy at application starts; and
\item[\textbf{(2,3)}] atomically updating both copies with a fine-grained concurrency control with helping.
\end{enumerate*}
We observe that the same invariant and concurrency control satisfy the three conditions and thus work also for our detectable version of SOFT.

%% file: reclamation.tex
\section{Safe Memory Reclamation}
\label{sec:reclamation}


All pointer-based lock-free DSs in DRAM should deal with the problem of safe memory reclamation (SMR).
For instance, Treiber's stack~\cite{treiber} is basically a linked list of elements with the head being the stack top, where a thread detaches the head block while another thread holds a local pointer to the same block.
Due to the local pointer, the reclamation of the detached block should be \emph{deferred} until every thread no longer holds such local pointers.
In most cases, the SMR problem is systematically handled with \emph{reclamation schemes} (think: lightweight garbage collection) such as hazard pointers (HP)~\cite{hp} and epoch-based reclamation (EBR)~\cite{ebr}.

Many of the prior works on persistent, pointer-based lock-free DSs in PM~\cite{} also handle the SMR problem with reclamation schemes.
Friedman \etal~\cite{friedman} use HP, and Li \etal{}~\cite{dss} and (nb)Montage~\cite{montage,nbmontage} use EBR.
However, we argue that the prior works do not properly discuss the subtlety arising from applying reclamation schemes to PM.
We address this subtlety by discovering and fixing a use-after-free bug of the prior works~(\cref{sec:reclamation:uaf});
preventing double-free error in case of a thread crash~(\cref{sec:reclamation:df}); and
preventing use-after-free error via the local pointers checkpointed in mementos~(\cref{sec:reclamation:memento}).
For a concrete context, we explain our fix and preventions on EBR.

\subsection{Fixing a Use-After-Free Bug}
\label{sec:reclamation:uaf}

\paragraph{Bug}

The queues of Friedman \etal{}~\cite{friedman} and Li \etal{}~\cite{dss} have a use-after-free bug caused by the lack of a flush before reclamation.
We explain the bug in the context of our algorithm for deletion in \cref{alg:smo:base} for presentation purposes.
If the location $loc$ were not flushed before retirement at L38, then it is possible that
\begin{enumerate*}
\item $old$ is retired at L38 and and reclaimed later;
\item the system crashes;
\item the CAS to $loc$ at L38 is not persisted and $loc$ still contains $old$ after the crash; and
\item a post-crash execution dereferences $old$.
\end{enumerate*}
The queues lack such a flush between a CAS and a retirement, possibly incurring a use-after-free error.

\paragraph{Fix}

A straightforward fix would insert a flush between the CAS and the retirement at L38, but we observe that such a flush is detrimental to the performance.
We mitigate the slowdown by exploiting the following property of EBR: if a memory block is retired in a \emph{critical section}, then it is never reclaimed in the same critical section.
Concretely, instead of flushing $loc$ at L38, we \emph{defer} the flush of $loc$ with a new \textsc{DeferFlush} API and actually perform such flushes in batch at the end of a critical section.
Batching is beneficial for two reasons:
\begin{enumerate*}
\item 
  we can merge multiple flushes to the same location; and
\item we can finish a critical section in an asynchronous manner so that the flushes are performed not in the critical path.
\end{enumerate*}
In our evaluation, we observe that the deferred flushes incur no noticeable runtime overhead.


\subsection{Preventing Double-Free in Thread Crashes}
\label{sec:reclamation:df}

\paragraph{Potential Error}

A straightforward application of SMR schemes may incur double-free error in case of a thread crash due to the thread's inability to detect whether a block is retired.
Suppose a thread crashes at L38 in \cref{alg:smo:base}, possibly before or after $\textsc{Retire}(old)$.
Then the post-crash recovery execution would go to L38 and perform CAS, \textsc{DeferFlush}, and \textsc{Retire} again.
It is safe to perform CAS and \textsc{DeferFlush} again because they are idempotent, but it is not the case for \textsc{Retire} as double retirements of $old$ incur a double-free error.

\paragraph{Prevention}

We prevent this error by
\begin{enumerate*}
\item retaining the critical section of a crashed thread and reviving it for the post-crash recovery execution; and
\item relaxing the retirement condition by allowing double retirements in a single critical section.
\end{enumerate*}
For the former, we propose a new API for retrieving the critical section by thread id; and
for the latter, we install a buffer of retired blocks and deduplicate it at the end of a critical section.




\subsection{Preventing Use-After-Free via Mementos}
\label{sec:reclamation:memento}

\paragraph{Potential Error}

For safe reclamation, we should clear local pointers not only in program variables but also in mementos.
Otherwise, it may be possible that blocks are reclaimed, the thread crashes, and then the operation retrieves and dereferences its memento's local pointers, invoking use-after-free error.

\paragraph{Prevention}

We prevent this error by clearing mementos just before the end of a critical section.
More specifically, we
\begin{enumerate*}
\item wrap a thread's operation on its root memento within a critical section; and
\item at the end of a critical section, invoke \textsc{Clear} on the root memento, which invalidates all PM locations checkpointed inside it.
\end{enumerate*}
For \textsc{Clear}'s crash consistency, we install a 1-bit ``clearing'' flag to the root memento, which is toggled and flushed at the beginning and the end of the method.
Should a crash happens, the monitor first checks whether the flag is set, and if so, resumes \textsc{Clear}.



%% file: evaluation.tex
\section{Evaluation}
\label{sec:evaluation}

We perform a performance evaluation of our programming framework for detectable DSs in PM.
For space purposes, we present only key results here and present the full evaluation results in \cite{supp}.
We use a machine running Ubuntu 20.04 and Linux 5.11 with dual-socket Intel Xeon Gold 6248R (3.0GHz, 24C/48T) and Intel Optane Persistent Memory 100 Series 256GB Module (DCPMM) configured in the App Direct mode.
We pin all threads to a single socket to ensure all DCPMM accesses are within the same NUMA node.

We have implemented our programming framework in Rust 1.59.0~\cite{rust} and built it with release mode.
We use Crossbeam~\cite{crossbeam} for memory reclamation and Ralloc~\cite{ralloc} for allocation in PM.
On top of the framework, we have implemented detectable versions the following DSs:
\textbf{\emph{MSQ-mmt-cas}}: Michael-Scott's queue (MSQ)~\cite{msq} with detectable CAS (\cref{sec:cas});
\textbf{\emph{MSQ-mmt-indel}}: MSQ with detectable insertion/deletion (\cref{sec:smo});
\textbf{\emph{MSQ-mmt-vol}}: MSQ with detectable insertion/deletion and DRAM scratchpad optimization (\cref{sec:volatile});
\textbf{\emph{CombQ-mmt}}: combining queue~\cite{pcomb};
\textbf{\emph{Clevel-mmt}}: Clevel hash table~\cite{clevel} with bug fixes~\cite{clevel-fix}; and
\textbf{\emph{SOFT-mmt}}: SOFT hash table~\cite{soft}.
We will compare the performance of our queues (\cref{sec:evaluation:queue}) and hash tables (\cref{sec:evaluation:hash}) with the prior works with and without detectability.

\subsection{Queue}
\label{sec:evaluation:queue}

\begin{figure*}[t]
  \centering
  \includesvg[width=2\columnwidth]{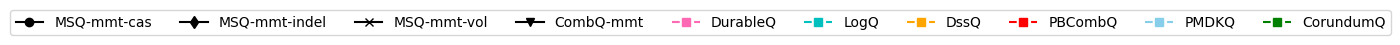}
  \begin{subfigure}[b]{0.5\columnwidth}
    \centering
    \includesvg[width=\columnwidth]{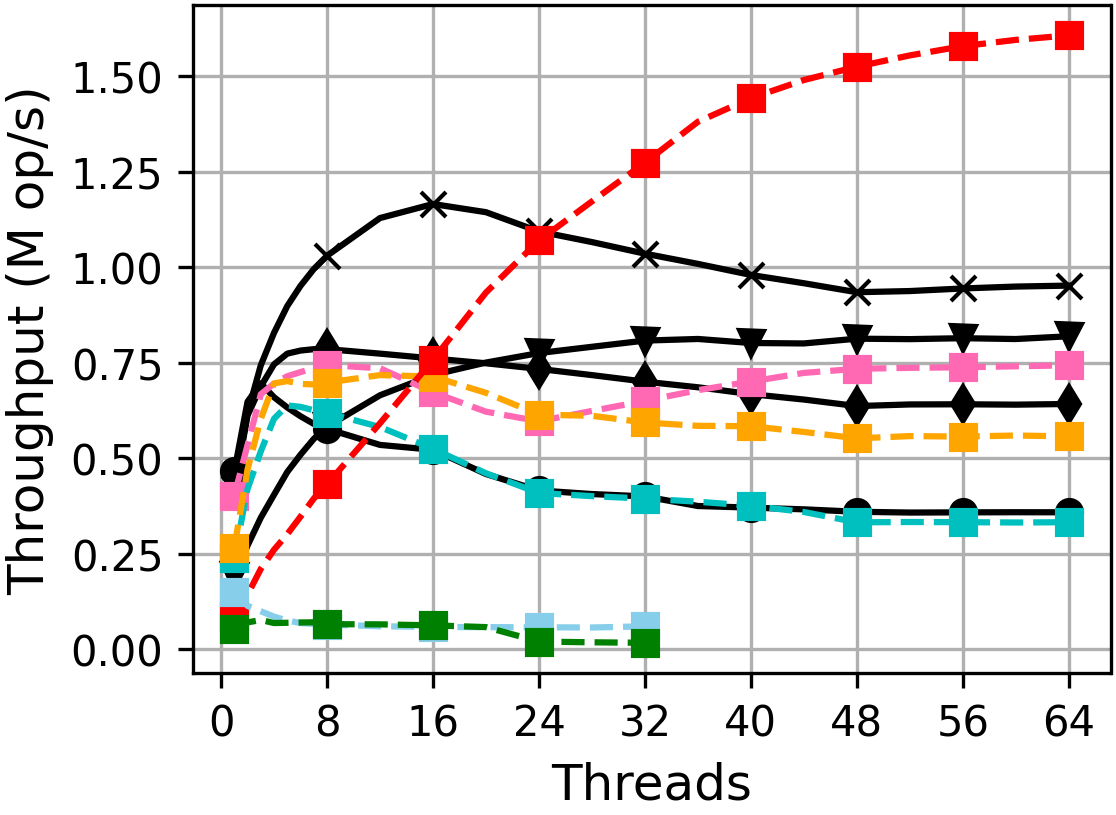}
    \caption{enqueue-dequeue pair}
    \label{fig:evaluation:queue-pair}
  \end{subfigure}
  \begin{subfigure}[b]{0.5\columnwidth}
    \centering
    \includesvg[width=\columnwidth]{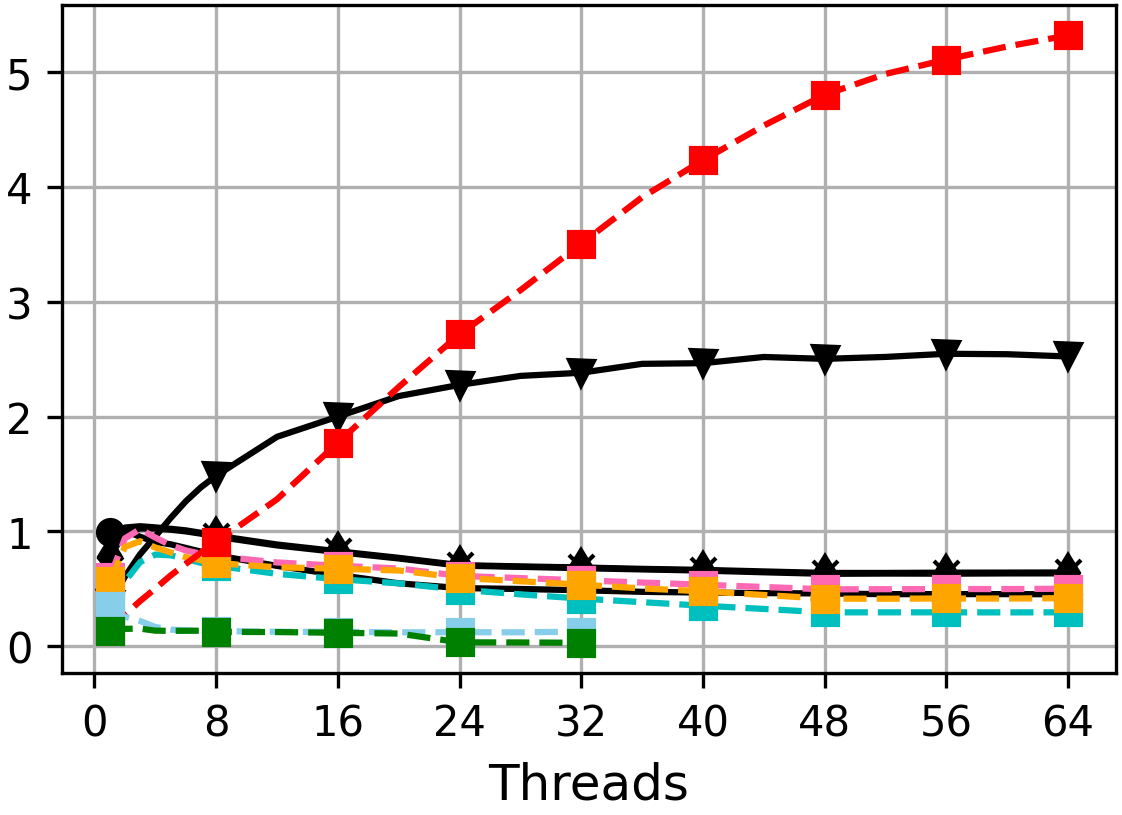}
    \caption{enqueue 20\%, dequeue 80\%}
    \label{fig:evaluation:queue-20}
  \end{subfigure}
  \begin{subfigure}[b]{0.5\columnwidth}
    \centering
    \includesvg[width=\columnwidth]{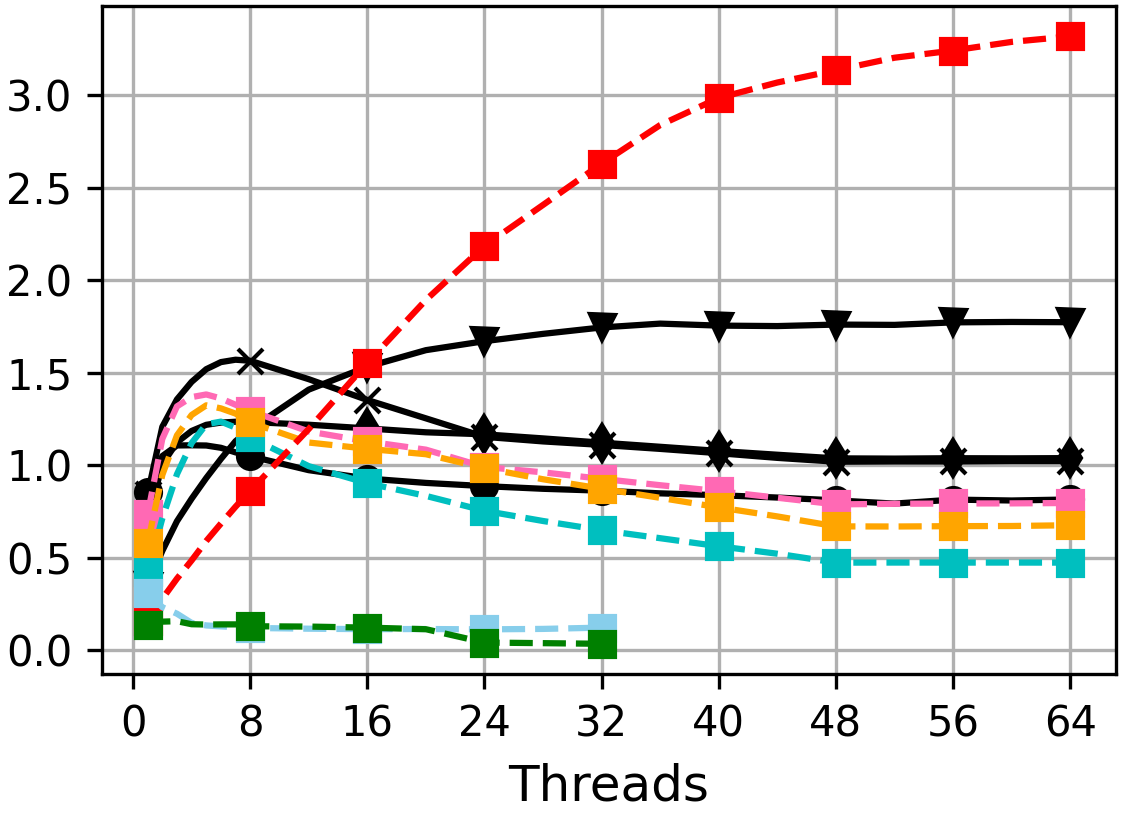}
    \caption{enqueue 50\%, dequeue 50\%}
    \label{fig:evaluation:queue-50}
  \end{subfigure}
  \begin{subfigure}[b]{0.5\columnwidth}
    \centering
    \includesvg[width=\columnwidth]{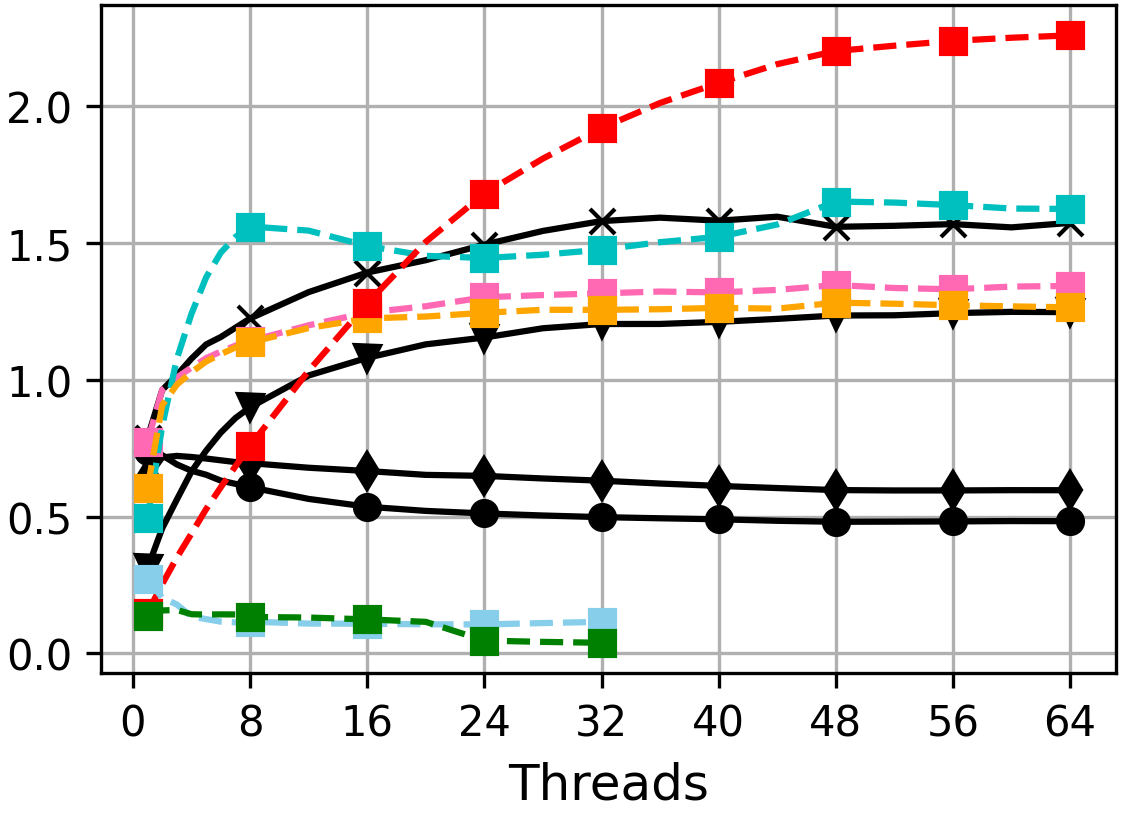}
    \caption{enqueue 80\%, dequeue 20\%}
    \label{fig:evaluation:queue-80}
  \end{subfigure}
  \vspace{-0.5em}
  \caption{Multi-threaded Throughput of Queues}
  \label{fig:evaluation:queue}
\end{figure*}

We compare the throughput of our queues with that of
\textbf{\emph{DurableQ}}: non-detectable durable MSQ by Friedman \etal{}~\cite{friedman};
\textbf{\emph{LogQ}}: detectable MSQ by Friedman \etal{}~\cite{friedman};
\textbf{\emph{DssQ}}: detectable MSQ by Li \etal{}~\cite{dss};
\textbf{\emph{PBcombQ}}: detectable combining queue by Fatourou \etal{}~\cite{pcomb};
\textbf{\emph{PMDKQ}}: non-detectable transaction-based queue in PMDK~\cite{pmdk}; and
\textbf{\emph{CorundumQ}}: non-detectable transaction-based queue in Corundum~\cite{corundum}.
We reimplement \emph{DurableQ} and \emph{LogQ} from scratch in Rust for fair comparison and a use-after-free bug~(\cref{sec:reclamation})
, and
implement \emph{DssQ} and \emph{PBcombQ} in Rust because their source code is not publicly available.
We observe the performance characteristic of \emph{PBcombQ} is different from \cite{pcomb}, because \emph{PBcombQ} fixes two linearizability bugs of \cite{pcomb} that we reported to the authors.


\cref{fig:evaluation:queue} illustrates the throughput of queues for four workloads:
\textbf{\emph{enqueue-dequeue}}: each operation enqueues an item and then immediately dequeues an item;
\textbf{\emph{enqueue-20\%}}: each operation enqueues (or dequeues) an item for the probability of 20\% (or 80\%);
\textbf{\emph{enqueue-50\%}}; and
\textbf{\emph{enqueue-80\%}}.
For each workload, we measure the throughput for a varying number of threads: 1 to 8 and the multiples of 4 from 12 to 64.
We report the average throughput for each scenario of 5 runs, each of which is executed for 10 seconds.
Before measuring the throughput, we populate the queues with 10M items to avoid too many empty dequeues.

We make the following observations.
\begin{enumerate*}
\item Transaction-based queues are noticeably slower than MSQs and combining queues.
  We omit the results of \emph{PMDKQ} and \emph{CorundumQ} for high thread counts because it took too much time to populate items.
\item Combining queues outperform MSQs for dequeue-heavy workloads.
  It is consistent with the observations made in \cite{pcomb}.
\item \emph{MSQ-mmt-vol} outperforms the other MSQs in PM with and without detectability for two reasons.
  \emph{First}, it outperforms \emph{DurableQ} because the former's dequeue writes the return value to mementos while the latter's dequeue writes to a newly allocated node.
  \emph{Second}, the timestamp-based checkpointing in mementos performs fewer PM flushes than the other detectable MSQs~(\cref{sec:overview:example}).
  The only exception is \emph{LogQ} for \emph{enqueue-80\%} with low thread counts, for which we are investigating the reason.
\item \emph{MSQ-mmt-cas} and \emph{MSQ-mmt-indel} perform comparably with the other MSQs in PM with and without detectability for dequeue-heavy workloads but not for enqueue-heavy workloads, because their enqueue perform CAS twice.
\item \emph{CombQ-mmt} incurs noticeable overhead over \emph{PBcombQ} because only the former supports reclamation.
  For safe reclamation, \emph{CombQ-mmt}'s combiner allocates a new memory block.
\end{enumerate*}

\subsection{Hash Table}
\label{sec:evaluation:hash}

\begin{figure*}[t]
  \centering
  \includesvg[width=1.5\columnwidth]{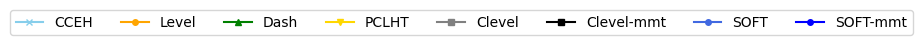}
  \begin{minipage}[b]{.45\textwidth}
    \includesvg[width=\columnwidth]{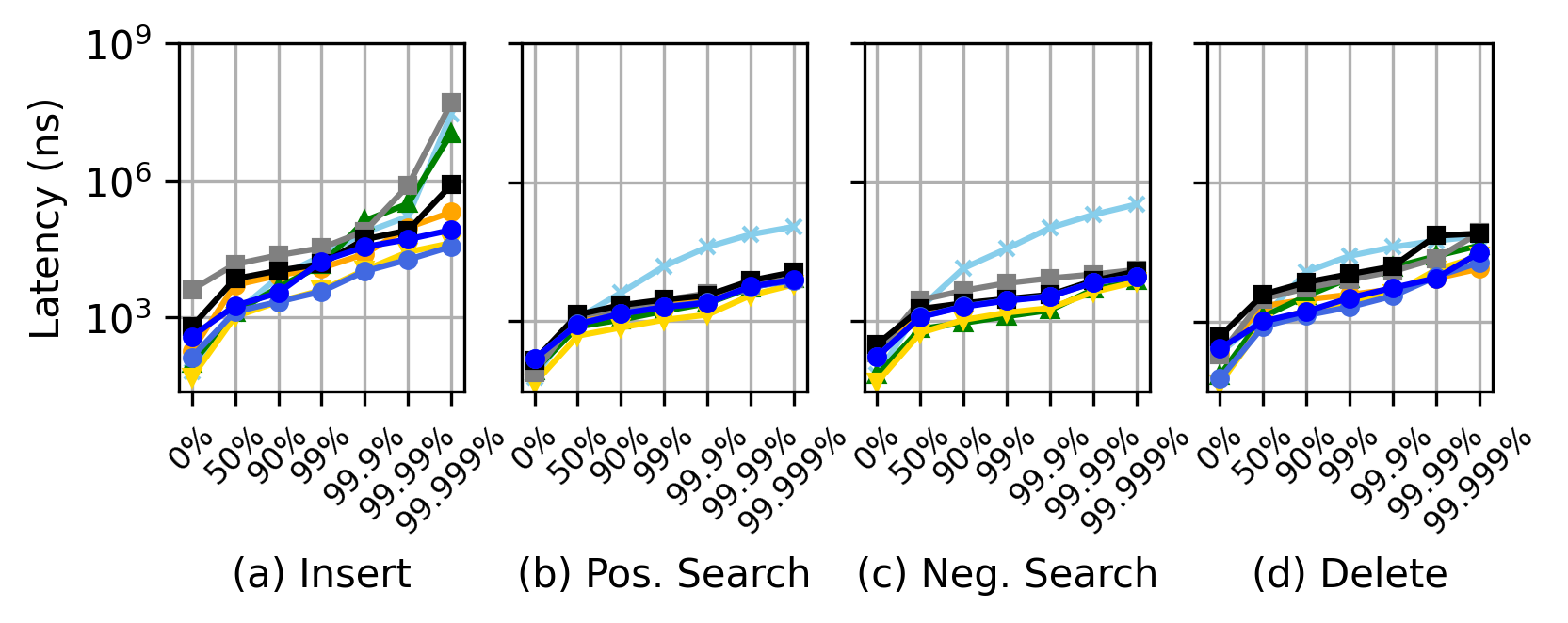}
    \caption{Tail Latency of Hash Tables with 32 Threads}
    \label{fig:evaluation:hash-latency}
  \end{minipage}
  \quad
  \begin{minipage}[b]{.45\textwidth}
    \includesvg[width=\columnwidth]{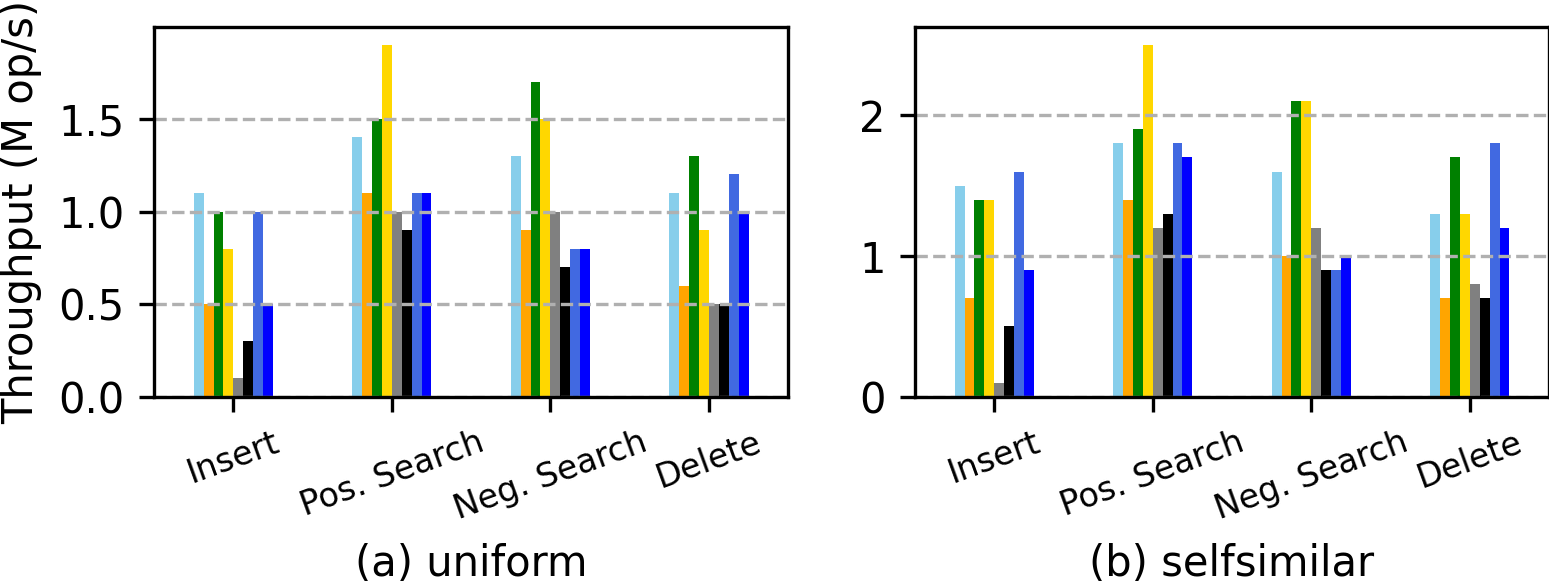}
    \caption{Single-threaded Throughput of Hash Tables}
    \label{fig:evaluation:hash-throughput-single}
  \end{minipage}
\end{figure*}

\begin{figure*}[t]
  \includesvg[width=2\columnwidth]{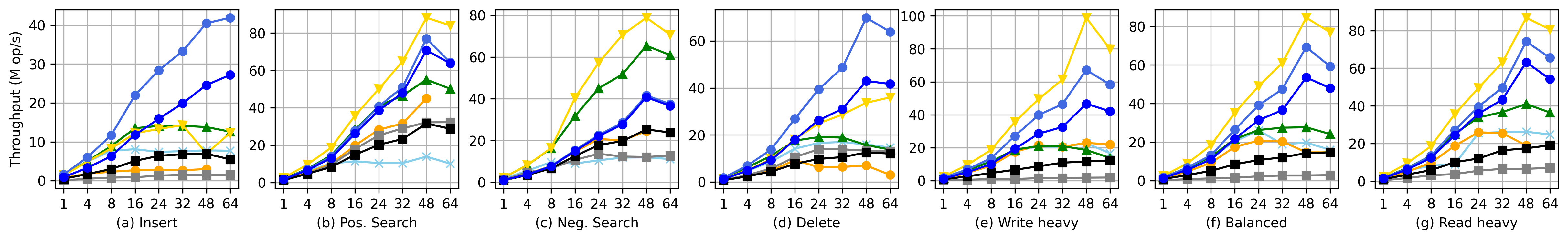}
  \caption{Multi-threaded Throughput of Hash Tables for Self Similar Distribution with Factor 0.2}
  \label{fig:evaluation:hash-throughput-multi-self-similar-0.2}
\end{figure*}

\begin{figure*}[t]
  \includesvg[width=2\columnwidth]{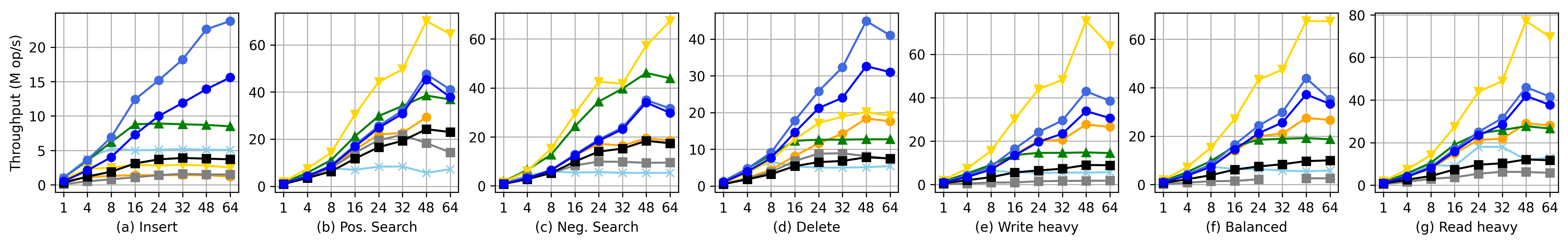}
  \caption{Multi-threaded Throughput of Hash Tables for Uniform Distribution}
  \label{fig:evaluation:hash-throughput-multi-uniform}
\end{figure*}


\noindent
We use the PiBench~\cite{pibench} benchmark because it is specifically designed for persistent hash tables.
We compare the throughput and latency of our hash tables with those of the persistent hash tables considered in an evaluation paper~\cite{hash-eval}:
\textbf{\emph{Clevel}}~\cite{clevel};
\textbf{\emph{SOFT}}~\cite{soft};
\textbf{\emph{CCEH}}~\cite{cceh};
\textbf{\emph{Dash}}~\cite{dash};
\textbf{\emph{Level}}~\cite{level}; and
\textbf{\emph{PCHLT}}~\cite{recipe}.
We use the implementation of these hash tables provided in \cite{hash-eval}.
Among them, we convert only lock-free ones according to \cite[Table 1]{hash-eval}, namely \emph{Clevel} and \emph{SOFT}, to detectable DSs.

\paragraph{Workloads}

We use exactly the same workloads as \cite{hash-eval}: ``we stress test each hash table with individual operations (insert, positive and negative search, and delete) and mixed workloads. Negative search means searching keys that do not exist.''
``We initialize hash tables with a capacity that can accommodate 16M key-value pairs.''
``To measure insert-only performance, we insert 200M records into an empty hash table directly.
To measure the performance of the search and delete operation and the mixed workloads, we first initialize the hash table with 200M items (loading phase), then execute 200M operations to perform the measurements (measuring phase).''
``We run the experiments with workloads using uniform distribution and skewed distribution (self similar with a factor of 0.2, which means 80\% of accesses focus on 20\% of keys). Since CCEH, Level hashing, and PCLHT have no support for variable-length keys and values, we consider fixed-length (8 bytes) keys and values.''

All our results show the same tendency as in \cite{hash-eval}, so we compare only \emph{Clevel-mmt} (and \emph{SOFT-mmt}) with \emph{Clevel} (and \emph{SOFT}, respectively).
For an evaluation of the other hash tables, we refer to \cite{hash-eval}.


\paragraph{Latency and Load Factor}

\cref{fig:evaluation:hash-latency} illustrates tail latency of hash tables for 32 threads under uniform distribution.
\emph{Clevel-mmt} (and \emph{SOFT-mmt}) shows similar tail latencies as \emph{Clevel} (and \emph{SOFT}, respectively) for all percentiles.

\emph{Clevel-mmt} shows almost the same load factor as \emph{Clevel}, because we did not change insert and resize schemes, and \emph{SOFT-mmt}'s and \emph{SOFT}'s load factors are always 100\% (see \cite{supp} for full results).

\paragraph{Single-threaded Throughput}

\cref{fig:evaluation:hash-throughput-single} illustrates single-threaded throughput of hash tables under uniform and skewed distributions.
For both distributions, we make the following observations.
\begin{enumerate*}
\item Detectable hash tables incur overhead over the original ones for insert and delete workloads (except for \emph{Clevel}'s insert, which we believe is due to implementation differences).
  It is expected from the fact that memento versions need to additionally checkpoint operation results.
\item Detectable hash tables show similar throughput with the original ones for search workloads.
  The reason is they execute almost the same instructions to read data.
\end{enumerate*}

\paragraph{Multi-threaded Throughput}

\cref{fig:evaluation:hash-throughput-multi-self-similar-0.2} and \cref{fig:evaluation:hash-throughput-multi-uniform} illustrate multi-threaded throughput of hash tables under uniform and skewed distributions, respectively.
We make the same observations as with single-threaded throughput.


%% file: conclusion.tex
\section{Related Work}
\label{sec:related}

\paragraph{Detectable Lock-Free DSs in PM}

Attiya \etal{}~\cite{nrl} presents a programming framework for detectable operations based on a novel detectable test-and-set (TAS) and CAS objects.
However, their TAS and CAS objects consume $O(P)$ and $O(P^2)$ spaces in PM, respectively, prohibiting its use for space-efficient DSs such as hash tables and trees; and
the authors present a few lock-free DS examples but do not present how to make lock-free DSs detectable in general.

Ben-David \etal{}~\cite{nrlp} presents such a general transformation with a novel concept of \emph{capsule}: each capsule serves as a unit of detectable operation, and at the end of a capsule execution, the program counter and local state is checkpointed and used as the starting point at recovery.
\begin{enumerate*}
\item While general, each capsule is required to follow one of the following forms to ensure its detectability: \emph{CAS-read}, \emph{Read-Only}, \emph{Single-Instruction}, and \emph{Normalized}~\cite{normalized}.
  In contrast, our transformation generally supports the SSA form~\cite{ssa1,ssa2}~(\cref{sec:overview:transformation}).
\item Capsule requires programmers to significantly restructure code for DRAM by explicitly delimiting programs into multiple capsules; and explicitly recovering states at the capsule boundaries.
\item Their detectable CAS object consumes $O(P)$ spaces in PM, while our detectable CAS object consumes 8 bytes for Intel Optane DCPMM.
\end{enumerate*}

Friedman \etal{}~\cite{friedman} and Li \etal{}~\cite{dss} present frameworks for detectable operations and detectable MSQs in PM, but both have a bug on reclamation~(\cref{sec:reclamation:uaf})
and perform slower than our MSQ due to an additional flush~(\cref{sec:evaluation:queue}).
Rusanovsky \etal{}~\cite{rusanovsky} and Fatourou \etal{}~\cite{pcomb} present hand-tuned detectable combining DSs, while ours is generally transformed from a volatile one.

\paragraph{Non-detectable Lock-Free DSs in PM}

Friedman \etal~\cite{friedman} presents non-detectable lock-free MSQs based on hazard pointers~\cite{hp} in PM.
Our detectable MSQ outperforms both of them because their dequeue operation allocates a new node to write the return value~(\cref{sec:evaluation:queue}).
Various hash tables~\cite{cceh,level,clevel,soft,dash,recipe} and trees~\cite{bztree,pactree} in PM have been proposed in the literature.
In this paper, we convert the Clevel~\cite{clevel} and SOFT~\cite{soft} hash tables to detectable DSs as case study because they are lock-free.
Converting the others to detectable DSs is an interesting future work (see \cref{sec:future} for details).

\paragraph{Transformation of DSs from DRAM to PM}

Izraelevitz \etal{}~\cite{izraelevitz} present a universal construction of lock-free DSs in PM, but it is reported that the constructed DSs are generally slow~\cite{nvtraverse,mirror}.
Lee \etal{}~\cite{recipe} propose a \emph{RECIPE} to convert indexes from DRAM to PM, but their guideline is abstract and high-level and not immediately applicable to DSs in general.
Kim \etal{}~\cite{pactree} propose the \emph{Packed Asynchronous Concurrency} (PAC) guideline to construct high-performance persistent DSs in PM, but their guideline is also abstract and high-level.
In contrast, our transformation is a more concrete guideline at the code level.

NVTraverse~\cite{nvtraverse} is a systematic transformation of a large class of persistent DSs exploiting an observation that most operations consists of two phases: read-only traversal and critical modification.
Then the traversal phase does not require flushes at all.
Mirror~\cite{mirror} is a more general and efficient transformation that replicates DSs in PM and DRAM scratchpad, significantly improving read performance.
FliT~\cite{flit} is a persistent DS library based on a transformation utilizing dirty cacheline tracking.
However, NVTraverse, Mirror, and FliT do not support transformation of detectable DSs.

\section{Future Work}
\label{sec:future}

Beyond those discussed in \cref{sec:related}, We will do the following future work.


\paragraph{Case Study}

Beyond lock-free DSs, we will perform more serious case studies with realistic storage engines to demonstrate our framework's scalability.
We are interested in designing detectable versions of traditional and distributed file systems, transaction processing systems for high-velocity real-time data~\cite{sstore}, and distributed stream processing systems~\cite{kafka} in our framework.
As a first step, we are currently building a lock-free PM file system.

\paragraph{Primitive Operation}

For such case studies, we expect we should design more primitive detectable operations for those used in realistic storage engines.
We are particularly interested in designing a detectable version of PMwCAS~\cite{pmwcas}, which is CAS for multiple words performed atomically.
PMwCAS is used in the BzTree persistent lock-free B-tree~\cite{bztree}, and we believe it can also be used in file systems and storage engines in general.
We are also interested in designing a detectable version of the standard constructs of fine-grained concurrency such as locks, combiners, and helpers.

\paragraph{Verification}

As an ongoing work, we are formally verifying the crash consistency and detectability of those DSs implemented in our framework in a program logic.
We believe that our framework has a great potential to lower the high cost of crash consistency verification because its stable nature unifies code for normal and recovery executions.
We will realize our framework's potential for lower-cost crash consistency verification by
\begin{enumerate*}
\item designing a separation logic for PM by recasting the idea of a rely-guarantee logic for PM \cite{persistent-rg} in the Iris separation logic framework~\cite{iris} on top of the Coq proof assistant~\cite{coq};
\item designing reasoning principles for detectable operations with mementos in the proposed separation logic; and
\item verifying detectable DSs.
\end{enumerate*}
